\documentclass[12pt]{article}

\usepackage{amssymb,amsthm,xspace}
\usepackage{graphicx}
\usepackage{amsmath}
\usepackage{cite}

\usepackage{color}
\usepackage{a4wide}

\newcommand{\C}{{\mathbb C}}
\newcommand{\R}{{\mathbb R}}

\newcommand{\im}{{\rm i }}
\newcommand{\ri}{{\rm i }}
\newcommand{\re}{{\rm e }}
\newcommand{\cV}{{\cal V}}

\newtheorem{theorem}{Theorem}

\def\tr{{\rm Tr\,}}
\def\det{{\rm det\,}}

\def\g{{h}}

\newcommand\be{\begin{eqnarray}}
\newcommand\ee{\end{eqnarray}}

\newcommand{\beq}{\begin{equation}}
\newcommand{\eeq}{\end{equation}}
\newcommand{\ber}{\begin{eqnarray}}
\newcommand{\eer}{\end{eqnarray}}

\begin{document}

\title{Anisotropic singularities \\ in chiral modified gravity}
\author{Yannick Herfray${}^{(a,\,b)}$, Kirill Krasnov${}^{(b)}$ and Yuri Shtanov${}^{(c)}$ \\ {}\\
{\it \small (a) Laboratoire de Physique, ENS de Lyon} \\
{\it \small 46, all\'{e}e d\'{I}talie, F-69364 Lyon Cedex 07, France}\\
{\it \small (b) School of Mathematical Sciences, University of Nottingham}\\ \it{\small University Park, Nottingham, NG7 2RD, UK}\\
{\it \small (c) Bogolyubov Institute for Theoretical Physics}\\ \it{\small 14-b
Metrologichna Str., Kiev, 03680, Ukraine}}

\date{v2: September 2016}
\maketitle

\begin{abstract}
In four space-time dimensions, there exists a special infinite-parameter family of chiral
modified gravity theories. All these theories describe just two propagating
polarizations of the graviton. General Relativity with an arbitrary cosmological constant
is the only parity-invariant member of this family.  We review how these modified gravity
theories arise within the framework of pure-connection formulation. We introduce a
new convenient parametrisation of this family of theories by using certain set of
auxiliary fields.  Modifications of General Relativity can be arranged so as to become
important in regions with large Weyl curvature, while the behaviour is indistinguishable from GR where Weyl curvature is small.  We show how the Kasner singularity of
General Relativity is resolved in a particular class of modified gravity theories of this
type, leading to solutions in which the fundamental connection field is regular all
through the space-time. There arises a new asymptotically De~Sitter region `behind' the
would-be singularity, the complete solution thus being of a bounce type.
\end{abstract}

\section{Introduction}

We seem to live in four space-time dimensions, and so should take the structures
available in this number of dimensions seriously. One of these is chirality. In the sense
used here, this is related to Lorentz group having two different types of spinor
representations. An object is called chiral if its spinor description uses more spinor
indices of one type than of the other. In particular, self-dual 2-forms are chiral
objects; see Appendix for a short review.

\subsection*{Chiral description of gravity}

A remarkable phenomenon occurs in General Relativity (GR) in four space-time dimensions.
This phenomenon is so stunning that we would like to refer to it as the {\em chiral
miracle\/}. It is well-known to experts. Still, even after almost 40 years since its
appearance in the literature, it has not become part of the background of all GR
practitioners. This is why we start our discussion with a review of this phenomenon.

The miracle is related to a decomposition of the Riemann curvature. In view of its
symmetries, the Riemann curvature $R_{\mu\nu\rho\sigma}$ can be thought of as a symmetric
$\Lambda^2\otimes \Lambda^2$-valued matrix, where $\Lambda^2$ denotes the space of
2-forms. In four dimensions, we can decompose 2-forms into their self-dual and
anti-self-dual parts $\Lambda^2 = \Lambda^+\oplus\Lambda^-$, and so we have the
decomposition of the Riemann curvature relative to the decomposition $\Lambda^2 =
\Lambda^+\oplus\Lambda^-$\,:
\beq\label{riemann}
{\rm Riemann} = \left( \begin{array}{cc}
A & B \smallskip \\
B^T & C \end{array} \right) \, .
\eeq
Here, $A$ and $C$ are symmetric $3\times 3$ matrices that can be referred to as the
self-dual--self-dual and anti-self-dual--anti-self-dual parts of the Riemann curvature,
respectively. For real metrics of Lorentzian signature, the matrix $C$ is the complex
conjugate of $A$, and the matrix $B$ is a Hermitian $3\times 3$ matrix $(B^T)^*=B$. The
Bianchi identity $R_{[\mu\nu\rho\sigma]}=0$ implies that the traces of $A$ and $C$ are
equal. In particular, for real Lorentzian metrics, this implies that the trace is real.

The above decomposition turns out to encode the irreducible pieces of the Riemann
curvature with respect to the Lorentz (or orthogonal in the case of Riemannian signature)
group. The trace of $A$ or $C$ is the scalar curvature. The trace-free parts of $A$ and
$C$ are the self-dual and anti-self-dual parts of the Weyl curvature, respectively. The
matrix $B$ encodes the trace-free part of the Ricci curvature.

Now comes the key point. In view of the above decomposition of the Riemann tensor, the
Einstein condition is equivalent to
\beq\label{einstein}
R_{\mu\nu} = \Lambda g_{\mu\nu} \qquad \Leftrightarrow \qquad \left\{
\begin{array}{rcl}
B &= &0 \, , \smallskip \\
{\rm Tr}\, A &= &\Lambda \, .
\end{array}
\right.
\eeq
This relation uses only the first row of the matrix in (\ref{riemann}). One could, of
course, use the second row equally well. Thus, {\em to impose the four-dimensional
Einstein condition, it is sufficient to have access just to half of the Riemann
curvature\/}. This is the chiral miracle occurring in four-dimensional GR.

It is sometimes erroneously thought that this phenomenon has to do with complexification.
Indeed, imposing some equation on a complex-valued quantity is equivalent to imposing two
equations on two real quantities --- the real and imaginary parts. However, this is not
what happens here. The best way to see this is to consider the Riemannian, or all-plus
signature. Most of the things stated above still apply, except that the two factors in
the decomposition $\Lambda^2=\Lambda^+\oplus \Lambda^-$ are no longer complex conjugates
of each other, both now being real (the Hodge star now satisfies $*^2=1$, with
eigenvalues $\pm 1$, so the eigenspaces are real). In this case, the matrices $A$, $B$,
and $C$ in (\ref{riemann}) are all real-valued, and $C$ is related to $A$ only by the
trace condition ${\rm Tr}\, A = {\rm Tr}\, C$ that still follows from the Bianchi
identity. The Einstein condition is still encoded in this case as given in
(\ref{einstein}). This makes it very clear that the fact that the Einstein condition is
encoded in half of the Riemann curvature has nothing to do with complexification, as it
still holds in the case of Riemannian signature, where the decomposition is real.

Reflection on the meaning of (\ref{einstein}) leads to the following important theorem:
\begin{theorem}[Atiyah--Hitchin--Singer '78]
Let $X$ be a 4-manifold with an Einstein metric. Then the induced connection on the
bundle $\Lambda^+$ of self-dual 2-forms is self-dual. Conversely, if the induced
connection on $\Lambda^+$ is self-dual, then the metric is Einstein.
\end{theorem}
Here, a connection is called self-dual if its curvature is self-dual as a 2-form. The
above theorem was discussed in \cite{Atiyah:1978wi} for metrics of Riemannian signature,
but it continues to hold equally well for Lorentzian metrics.

The above theorem is the basis of what can be called chiral descriptions of the
four-dimensional GR\@. There are several related versions of these, to be reviewed in the
main text. The main idea is that some chiral (hence, complex-valued in the case of
Lorentzian signature) object is used for the description of geometry, instead of the
real-valued non-chiral metric on which the usual description is based. As can be
anticipated from the fact that only half of the Riemann curvature is needed, such chiral
descriptions are more economical than the metric GR, which justifies interest in them.

\subsection*{Chiral modifications of GR}

While the chiral miracle described above is certainly known to differential geometers
specialising in Einstein manifolds, there is another related miracle that is almost
completely unknown to the community. It is the fact that the four-dimensional Einstein
condition can be non-trivially deformed in a chiral way.

It is well known that GR can be modified, the simplest example of such a modification
being the $R^2$ gravity, of relevance, e.g., as a good model of inflation
\cite{Starobinsky:1980te, Ade:2015lrj}. This model is equivalent to GR coupled to an
additional scalar field, and so propagates not just the two polarisations of the graviton
as in GR, but also a scalar. One can consider more involved modifications of GR with
higher powers of the curvature added to the Lagrangian. One can quickly convince oneself
that, because of the higher derivatives present in these modified theories, they all
propagate more degrees of freedom (DOF) than does GR\@. If one insists on second-order
field equations, then GR is the unique theory, at least in four dimensions. It thus seems
impossible to modify GR without adding extra propagating DOF to it. This is the content
of several GR uniqueness theorems available in the literature.

It then comes as a big surprise that it is indeed possible to modify GR without adding
extra DOF if one starts from one of its chiral descriptions. The resulting chiral
modified gravity theories continue to have second-order field equations. A complete count
of the number of degrees of freedom by the Hamiltonian analysis shows that they have the
same number of propagating degrees of freedom as in GR
--- the two propagating polarisation of the graviton. And, as will be reviewed below, there is
an {\em infinite-parametric\/} class of such chiral modified gravity theories, in which
GR is just a special member.

Once GR gets embedded into an infinitely large class of gravity theories all with similar
properties, one is forced to ask a number of questions: What makes GR unique as compared
to all these other theories? Why is the real world described so well by GR? Is it
described by GR perfectly, or is this only an approximate truth? To put it differently,
the very fact that such chiral modified gravity theories exist makes one obliged to
understand them.

One, however, faces the following difficulty.
In the chiral description of GR with Lorentzian signature, the metric ceases to be a
fundamental field variable. One rather deals with a
complex-valued field subject to appropriate `reality conditions' to ensure the
reality of the metric that this field defines. In GR, this procedure is (reasonably) well understood.
In the case of chiral modifications of GR, the fundamental complex-valued
field and the general structure of the theory remain intact; only the field dynamics gets
modified continuously without appearance of new degrees of freedom. What remains much
less understood, however, is how one is to modify the `reality conditions.' Note that,
since the metric is a derived field in the theory, it may no longer be required to
remain real-valued in all cases in a chiral modification of GR\@. It appears that one
cannot establish proper general `reality conditions' before one considers coupling of
gravity with matter, --- the problem that lies beyond the scope of the present paper and
still awaits for its solution.

%Having asked this question, we also have to explain one possible reason why this
%question, as well as these chiral modifications of GR, can be ignored. This has to do
%with the complex-valued nature of chiral objects in the case of Lorentzian
%metric signature. Thus, in the Lorentzian case, one has to
%deal with complex-valued objects subject to appropriate reality conditions. For the case
%of chiral description of GR this procedure is (reasonably) well understood.
%For the case of the chiral modified theories, this is understood much less. It moreover
%appears that in general it is even inconsistent to demand that the space-time metric is
%real. Thus, there are some examples where the condition that the space-time metric is
%real seems to be incompatible with the dynamics of the theory. The situation here can be
%compared to some system in classical mechanics where there is a term explicitly
%containing the imaginary unit in the Lagrangian. In some cases it is possible to make
%sense of such a system by imposing appropriate reality conditions. But there are also
%cases when this may be impossible. The difficulties facing the chiral modified gravity
%theories are of this sort. Thus, at present, it is not clear if these modified theories
%admit any physical interpretation. It is on these grounds that they can be ignored.

Having said that, we should also remark that there are many situations where the chiral
modified gravity theories behave perfectly sensibly and admit the usual interpretation in
terms of a real-valued space-time metric. Moreover, one can also formulate these modified
theories as theories of Riemannian signature metrics. In this case, all objects involved
are real-valued, and the issues referred to above simply do not arise.

We should also stress that the type of modifications of gravity we are interested in here
is unique in the following sense. One can inspect the proofs of the GR uniqueness,
notably the modern proofs that deal with the scattering amplitudes, and note the
particular assumptions in those proofs that are violated by our chiral theories. Removing
those assumptions, one can see that there results a new `uniqueness' theorem stating that
these chiral modifications are the only ones that describe propagating gravitons with
second-order field equations; see \cite{Krasnov:2014eza} for an argument of this sort.

All this encourages us to investigate these chiral modifications of GR in detail.

\subsection*{This paper}

In this paper, as a step towards a better understanding of these modified gravity
theories, we study them in a particular setup of Bianchi~I space-times. Already in the
case of GR, this is a rich setup, exhibiting the famous Kasner singular behaviour. The
Kasner behaviour is widely expected to capture the essence of a generic spacelike
singularity of General Relativity.

Our present work is related to earlier works by some of us \cite{Krasnov:2007uu,
Krasnov:2007ky}  (see also \cite{Krasnov:2008sb}), where the spherically symmetric
problem was analysed. It was found that there is a peculiar mechanism that resolves the
singularity inside a black hole in terms of the fundamental connection fields.
Homogeneous and isotropic cosmological solutions in the modified gravity theory
under investigation coincide with those of general relativity. The effects of
modifications on the theory of cosmological perturbations were under investigation in
\cite{Krasnov:2010tt}.

Previous works on the physics of these modified gravity theories used a first-order
formulation. The corresponding computations involved many auxiliary fields and were quite
cumbersome. One of the aims of this paper is to put to use the recently developed and
much more economical pure-connection formulation. As compared to previous works on the
subject, this simplifies the analysis tremendously.

Our main result is that a generic class of modified gravity theories of the type studied
here resolves the Kasner singularity of the GR solution of the Bianchi~I model. This
resolution occurs in a way similar to the case of black hole mentioned above: although
the metric field based on this solution still contains singularities and
experiences changes of signature, the fundamental connection field is everywhere regular.

To avoid misunderstanding, we would like to stress that in the family of modified gravity theories that we study modifications can be arranged so as to become important only in regions of large Weyl curvature. In the Bianchi I setup that we study this is the region near the would-be singularity. In regions of small Weyl curvature the modifications are negligible and these gravity theories behave indistinguishable from GR. To put it differently, the chiral modified gravity theories are continuously deformable to GR. So, by choosing the parameter(s) controlling the modification to be sufficiently small the modified theories will behave as GR on large scales, while exhibit behaviour very different from GR at small distances. The scale where modification becomes important can be chosen to be close to the Planck scale, and so one can view the setup that we study as a classical theory of gravity where gravity gets strongly modified close to the Planck scale, while remaining indistinguishable from GR at large distances. 

\subsection*{New description of chiral modified gravity theories}

Apart from direct application to Bianchi~I models, this paper develops a new
parametrisation of the underlying modified gravity theories. The new description uses not
just a connection, but also a number of auxiliary fields assembled into a $3\times 3$
matrix. We refer to a description with auxiliary fields as `mixed' in the main text,
because it is half-way between the first-order formulation of the theory with even more
fields, and the pure-connection formulation. The new mixed description turns out to be
quite powerful. In particular, it allows for a description of GR with zero cosmological
constant in an essentially pure-connection setting, something which is impossible if one
works just with connections.

Although the new `mixed' description of our modified gravity theories has been originally
developed specifically for the present analysis of the Bianchi~I setup, it later proved
to be very powerful for a variety of other problems. Two recent papers
\cite{Fine:2015hef} and \cite{Krasnov:2015qhm}, devoted to different issues, are
substantially based upon this `mixed' description.

\subsection*{Organisation of the paper}

In Section \ref{sec:pure-conn}, we describe the so-called pure-connection formulations of
gravity. Modifications of gravity of the type studied in this paper are easiest to
introduce in this pure-connection setup. We start with the Plebanski first-order
description of GR\@. We then show how the chiral pure-connection formulation of GR
arises, and motivate the modified theory as its certain generalisation.  In this section,
we also present the `mixed' new parametrisation of the modified theory. In Section
\ref{sec:Bianchi}, we specialise to the sector of interest, which is that of Bianchi~I
connections. We introduce the evolution equations for the connection components and
obtain the metric described by this connection. In Section \ref{sec:general}, a
convenient choice of the time variable is made, which allows us to solve the evolution
equations for our theories in full generality, without making any assumption about the
function that controls the modification. We also establish some general properties of the
solution. Section \ref{sec:GR} specialises to the case of GR with non-zero cosmological
constant. We describe the connection components in this case, and obtain the Kasner
behaviour of the metric, which is realised near the singularity. In Section
\ref{sec:modified}, we analyse the solution in the case of a particular one-parameter
family of modifications. It is here that we will see how the modification resolves the
Kasner singularity. We end up with a discussion.

There is a number of appendices at the end of the paper. In the first Appendix, we review
some basic facts about spinors and self-duality in four dimensions. The second Appendix
starts with a more general assumption about the connections of interest, and derives the
ansatz used in the main text. In the third Appendix, for completeness, we derive the GR
solution working in physical (proper) time. This is possible, but is considerably more
involved than the derivation presented in the main text. In the last Appendix, we derive
the Bianchi~I solution in GR with $\Lambda = 0$ working in the `mixed' parametrisation
with auxiliary fields.

\section{The pure-connection formulation(s) of gravity}
\label{sec:pure-conn}

One of the simplest ways to understand modifications of gravity of our type is to first
formulate GR as a theory of connections. We will describe how this is done below. For
now, let us just say that the connection plays the role of the potential for the metric,
schematically\footnote{This and subsequent schematic equations encode, in general,
non-linear relations between right-hand and left-hand sides.} $g=\partial A$, so that the
metric is constructed from the first derivative of the connection. The gravity Lagrangian
is constructed as an algebraic function of the curvature of the connection. A particular
such function gives dynamics equivalent to that in GR, see below.

In the pure-connection formulation, GR can be straightforwardly modified by considering
an arbitrary function of the curvature as the Lagrangian. This is guaranteed not to
change the second-order character of field equations. Sometimes it also does not change
the number of propagating degrees of freedom of the theory; see below.

At the very basic level, the way this approach works is as follows. There is some action
principle with the only field appearing in the Lagrangian being the connection $A$. The
arising Euler--Lagrange equations are second order partial differential equations for the
connection, which schematically can be expressed as $\partial^2 A=A$.  For an arbitrary
connection, not necessarily satisfying its field equations, one can construct a certain
metric $g_A$ from (the first derivatives of) $A$. The Ricci tensor of $g_A$ is then of
third order in derivatives of $A$. By using the connection field equations, $\partial^2 A
= A$, the Ricci tensor gets converted into an object of first order in derivatives of
$A$. For the case of GR in this framework, this is just a multiple of the metric itself,
and we obtain the vacuum Einstein equations. Thus, schematically,
\beq
{\rm Ricci}[g_A] = \partial^2 g_A = \partial^3 A \underset{\hbox{\small on-shell}}{=}
\partial A = g_A \, .
\eeq
This explains how the field equations for the connection imply the Einstein equations for
the metric $g_A$. Another similar line of identities can be written for the Weyl
curvature of the metric $g_A$. In this case, the connection field equations imply that
this Weyl curvature becomes (on-shell) identical to the curvature $F[A]$ of the
connection itself:
\beq
{\rm Weyl}[g_A] = \partial^2 g_A = \partial^3 A \underset{\hbox{\small on-shell}}{=}
\partial A = F[A] \, .
\eeq

The on-shell identification of the Weyl curvature with the curvature of $A$ makes it
clear that an interesting class of modifications of gravity can be obtained in this
formulation by simply appending the Lagrangian of GR with various scalars constructed
from $F[A]$. These modifications will become important in regions where the Weyl
curvature is large, and will thus have a chance to change the qualitative behaviour of
solutions near anisotropic singularities. At the same time, such modifications of the
Lagrangian for gravity do not lead to an increase in the order of field differential
equations, which remain to be second-order partial differential equations for $A$.

The above discussion makes it clear that this approach to modifying GR is available in
any `pure-connection' formulation. There are several possible such formulations, as we
shall now review. However, the above argument only guarantees that the order of field
differential equations is unchanged, but it does not guarantee that the number of
propagating degrees of freedom remains the same after modification. And indeed, the
argument with scattering amplitudes \cite{Krasnov:2014eza} suggests that most of such
modifications will introduce additional degrees of freedom.

After a brief review of the known pure-connection formulations, we will focus on the
chiral pure-connection formulation, which is capable of modifying gravity without
introducing new degrees of freedom.

\subsection{Eddington--Schr\"{o}dinger formulation}

This is the pure-connection formulation of GR that is as old as the Einstein--Hilbert
theory itself. As with all pure-connection formulations, a particularly simple way to get
it is to start with the first-order formulation in which one has both the connection and
the metric as independent variables. In the case of GR, this is the Palatini formulation,
with the affine connection $\Gamma_{\mu\nu}{}^\rho$ and the metric $g_{\mu\nu}$. One then
`integrates out' the metric $g_{\mu\nu}$, i.e., solves the field equations for $g$ and
substitutes the solution back into the action. This is only possible to do in the
presence of a non-zero cosmological constant. As a result, an action principle is
obtained that contains only $\Gamma_{\mu\nu}{}^\rho$. In the Eddington--Schr\"{o}dinger
case it is constructed from the (symmetric part of the) Ricci tensor
\beq\label{Ricci-Gamma}
R_{\mu\nu}[\Gamma] := R_{(\mu \alpha \nu)}{}^\alpha[\Gamma]\, ,
\eeq
where $R_{\mu\nu\rho}{}^\sigma[\Gamma]$ is the curvature of $\Gamma_{\mu\nu}{}^\rho$, the
latter being assumed to be symmetric in its lower two indices. The pure-connection action
principle is then
\beq\label{Eddington}
S[\Gamma] = \frac{1}{8\pi G\Lambda} \int \sqrt{ \det  \left( R_{\mu\nu}[\Gamma]
\right)}\, d^4 x \, .
\eeq
It is not hard to check that the square root of the determinant of $R_{\mu\nu}[\Gamma]$
has the correct density weight so that the integral over the spacetime is well-defined.

The field equations that result from (\ref{Eddington}) are
\beq\label{E-feqs}
\nabla_\rho R^{\mu\nu}[\Gamma] = 0\, ,
\eeq
where $R^{\mu\nu}[\Gamma]$ is the inverse of $R_{\mu\nu}[\Gamma]$. If one now makes a definition
\beq \label{g-gam}
g_\Gamma:= \frac{1}{\Lambda} R_{\mu\nu}[\Gamma]\, ,
\eeq
then equation (\ref{E-feqs}) tells that $\Gamma_{\mu\nu}{}^\rho$ is the
$g_\Gamma$-compatible connection. The definition (\ref{g-gam}) of $g_\Gamma$ is then the
vacuum Einstein equation.

The above description shows that this pure-connection formulation only makes sense when
$\Lambda\not=0$. Another feature of this formulation is that the field equations
(\ref{E-feqs}) are highly non-polynomial in the derivatives of $\Gamma$, containing the
inverse of $R_{\mu\nu}[\Gamma]$. Both these features are shared by any pure-connection
description of gravity.

Theory (\ref{Eddington}) can be modified by considering more general functionals of
$\Gamma$. For example, one can imagine dropping the symmetrization in (\ref{Ricci-Gamma})
while keeping the same action (\ref{Eddington}). One can also take another contraction
$R_{\mu\nu\alpha}{}^\alpha$ and form the same Eddington functional from it. More
generally, any scalar density of weight one constructed from the Riemann tensor
$R_{\mu\nu\rho}{}^\sigma(\Gamma)$ can serve as Lagrangian. It would be interesting to
classify the freedom available here, and also to study the arising modifications. As we
have already indicated above, it should be expected that modifications of the Eddington
formalism will describe more degrees of freedom compared to those in GR\@.

\subsection{The spin-connection formulation}

To get this formulation, one proceeds in a similar way starting from a different
first-order description. We can take it to be the Einstein--Cartan tetrad formulation.
The trick of integrating out the vielbein from the first-order formulation can be carried
out in $(2+1)$ dimensions \cite{Peldan:1991mh}. This is best explained in Section 3.4 of
\cite{Peldan:1993hi}. Again, the nonzero cosmological constant is essential, and again
one obtains a Lagrangian built from the square root of the determinant of the matrix of
curvatures, which is characteristic of these approaches.

We are not aware of any pure spin-connection formulation in four spacetime dimensions. To
obtain such a formulation, one would need to integrate out the tetrad, which in this case
amounts to solving the equation
\beq\label{EC-t}
\epsilon^{IJKL} \theta^J\wedge F^{KL}[\omega] = \Lambda \epsilon^{IJKL} \theta^J\wedge
\theta^K\wedge \theta^L\, ,
\eeq
where $\omega^{IJ}$ is the spin connection, $\theta^I$ is the tetrad, and the capital
indices are the `internal' four-dimensional ones. One would need to solve the above
equation for the tetrad to obtain an algebraic function of the curvature
$F^{IJ}_{\mu\nu}$. The pure-connection Lagrangian is then the determinant of the
resulting tetrad. In contrast to all cases already considered where the solution presents
itself readily, equation (\ref{EC-t}) is very non-trivial to solve, although it is not
impossible that the corresponding pure-connection formulation does exist. It would be
interesting to find it.

After the first version of the present manuscript has appeared on the arXiv, an
interesting paper \cite{Basile:2015jjd} appeared. In particular, this paper shows how
(\ref{EC-t}) can be solved perturbatively, in terms of an expansion around a given
background tetrad. The resulting pure-connection action has also been computed.

Another line of attack on this problem can be to replace the object
$\theta^I\wedge\theta^J$ in the Einstein--Cartan Lagrangian by a 2-form object $B^{IJ}$.
One can then add a term with a Lagrange multiplier field imposing the constraint that
$B^{IJ}$ is of the type $\theta^I\wedge\theta^J$; see \cite{DePietri:1998hnx}. After
that, one can integrate out $B^{IJ}$ as well as the Lagrange multiplier field, as we do
later in the case of the chiral formulation. It would be interesting to perform this
exercise and to compare the result with that of \cite{Basile:2015jjd}.

\subsection{The chiral Plebanski formulation}

In the chiral approach, one starts with the Plebanski formulation of GR
\cite{Plebanski:1977zz} (see also \cite{Capovilla:1991qb, Krasnov:2009pu}). This is a
first-order formulation, with an ${\mathfrak su}(2)$ Lie-algebra-valued two-form field
and a connection as independent variables. This formulation gives a concrete Lagrangian
realisation of the Atiyah--Hitchin--Singer theorem reviewed in the Introduction. Let us
start by describing this formulation, which is going to be the basis for all
constructions in the present paper.

We denote the Lie-algebra indices by lower-case Latin letters $i,j,k,\ldots =1,2,3$. The
basic fields are a Lie-algebra-valued two-form field with components $B^i$ and a
connection one-form with components $A^i$. There is also a Lagrange multiplier field
$\Psi^{ij}$, which is symmetric and traceless. The action of the theory is
\beq\label{plebanski}
S[B,A,\Psi] = \im \int B^i \wedge F^i - \frac{1}{2} \left(\Psi^{ij} +
\frac{\Lambda}{3}\delta^{ij}\right) B^i \wedge B^j \, .
\eeq
Here, $\Lambda$ is the cosmological constant, which may be zero. The imaginary unit
$\im=\sqrt{-1}$ in front of the action is needed in order to make it real for fields
satisfying the reality conditions as appropriate for Lorentzian signature; see below. One
would not need this pre-factor in either Riemannian or split signature.

Let us consider the field equations stemming from (\ref{plebanski}). First, varying with
respect to $B^i$, we get
\beq\label{Pleb-main-eqn}
F^i = \left(\Psi^{ij} + \frac{\Lambda}{3}\delta^{ij}\right) B^j \, .
\eeq
The Euler--Lagrange equation for the connection is
\beq\label{compat}
d_A B^i = 0 \, .
\eeq
Finally, there is the equation obtained by varying with respect to $\Psi^{ij}$\,:
\beq\label{simpl}
B^i \wedge B^j \sim \delta^{ij}\, .
\eeq
This equation can be understood as telling that $B^i$ `come from a tetrad' in the sense
that $B^i$ satisfying this equation contain no more information than that provided by the
metric plus a choice of an ${\rm SO}(3)$ frame at every spacetime point. Equation
(\ref{compat}) can be solved for $A^i$ in terms of derivatives of $B^i$ whenever $B^i$
are non-degenerate (that is, the three two-forms $B^i$ are linearly independent). In
particular, equation (\ref{compat}) can be solved if the two-forms $B^i$ satisfy
(\ref{simpl}), in which case the solution $A^i$ can be shown to be just the self-dual
part of the Levi-Civita connection for the metric described by $B^i$. Equation
(\ref{Pleb-main-eqn}) then becomes a statement that the curvature of the self-dual part
of the Levi-Civita connection of a metric is self-dual as a two-form.
Atiyah-Hithich-Singer theorem stated in the Introduction guarantees this to be equivalent
to the Einstein condition, which shows that (\ref{plebanski}) is indeed a description of
GR\@. When all field equations are satisfied, i.e., on-shell, the field $\Psi^{ij}$ is
identified with the self-dual part of the Weyl curvature. For more details on this
formulation, the reader is referred, e.g., to \cite{Krasnov:2009pu}.

\subsection{The metric}

In the above description, we have mentioned the fact that $B^i$ satisfying (\ref{simpl})
contain no more information than that available in the metric (up to gauge rotations).
This metric is determined by the two-form fields directly, as we now review.

The main point is the geometric idea that, in four dimensions, one knows the conformal
class of the metric (i.e., the metric modulo multiplication by an arbitrary function) if
one knows which two-forms are self-dual. Therefore, by declaring the triple of two-forms
$B^i$ to be self-dual, one fixes the conformal class of the metric uniquely. Explicitly,
the conformal class is given by the following representative due to Urbantke
\cite{Urbantke:1984eb}:
\beq\label{urb}
g_{\mu\nu} \sqrt{\det  g} \sim \tilde{\epsilon}^{\alpha\beta\gamma\delta} \epsilon^{ijk}
B^i_{\mu\alpha} B^j_{\nu\beta} B^k_{\gamma\delta}\, .
\eeq
Here $\tilde{\epsilon}^{\alpha\beta\gamma\delta}$ is the anti-symmetric tensor density of
weight one which in any coordinate system has components $\pm 1$, and the proportionality
means equality up to an arbitrary positive coordinate-dependent scalar factor. To fix the
metric completely, it suffices to fix this factor, or to specify the associated volume
form. In the case of the Plebanski theory that described GR, the associated volume form
${\mathcal V}$ is fixed uniquely from the requirement that the metric be Einstein. The
correct volume form then turns out to be:
\beq\label{vol-BB}
3!\, \im\, {\mathcal V} = B^i\wedge B^i \, .
\eeq

The content of the previous subsection can then be summarised by saying that when field
equations (\ref{Pleb-main-eqn})--(\ref{simpl}) are satisfied, the metric determined by
(\ref{urb}) with the associated volume form determined by (\ref{vol-BB}) is Einstein.

The above discussion applies to complexified GR, in which case all fields in the
Plebanski Lagrangian are considered to be complex-valued. Different so-called `reality
conditions' are to be imposed to get different metric signatures. The easiest choice is
the split signature, for which one simply takes all fields to be real. In terms of the
gauge group, this corresponds to the real form ${\rm SL}(2,\R)$ of ${\rm SL}(2,\C)$. The
Riemannian signature is obtained by taking the real form ${\rm SU}(2)$. Finally, to get
the metrics of Lorentzian signature, one needs to work with complex-valued fields but
impose the reality conditions (here and below, an overbar denotes complex conjugation)
\beq\label{reality-B}
B^i \wedge \bar B^j = 0
\eeq
together with the condition of reality of the volume form (\ref{vol-BB}). The reality
conditions (\ref{reality-B}) guarantee that the complex conjugate forms $\bar B^i$ are
wedge product orthogonal to $B^i$. Given that we want to identify $B^i$ with self-dual
two-forms, and self-dual two-forms are wedge product orthogonal to anti-self-dual ones,
equation (\ref{reality-B}) is the correct reality condition because it simply states that
anti-self-dual two-forms are complex conjugates of self-dual ones. In other words, the
reality condition (\ref{reality-B}) guarantees that the conformal metric determined by
(\ref{urb}) is real Lorentzian.

\subsection{Modified Gravity}

A particular family of modified gravity theories, inspired by the Plebanski formulation,
was motivated and described in \cite{Krasnov:2006du}. In one possible parametrisation,
the modification is to allow the cosmological constant in (\ref{plebanski}) to be an
arbitrary ${\rm SO}(3)$-invariant function of the field $\Psi^{ij}$\,:
\beq\label{pleb-modified}
S[B,A,\Psi] = \im \int B^i \wedge F^i - \frac{1}{2} \left(\Psi^{ij} +
\frac{\Lambda(\Psi)}{3}\delta^{ij}\right) B^i \wedge B^j \, .
\eeq
It can be shown \cite{Krasnov:2008zz} by the Hamiltonian analysis that this theory
continues to propagate just two degrees of freedom, similarly to GR\@. At the same time,
this is a modified theory of gravity, in which modification becomes important in
spacetime regions where the function $\Lambda(\Psi)$ significantly deviates from a
constant. A particularly simple one-parameter family of modifications is obtained by
considering the function $\Lambda(\Psi)$ in the form of a quadratic polynomial in
$\Psi^{ij}$\,:
\beq\label{alpha-family}
\Lambda(\Psi) = \Lambda_0 - \frac{\alpha}{2} \tr (\Psi^2) \, ,
\eeq
where $\alpha$ is an arbitrary parameter with dimensions $L^2$ (i.e., inverse curvature).
For this family of modified theories, one expects strong deviations from GR when the
field $\Psi$ (encoding the Weyl curvature in GR) becomes of the order of
$1/\alpha$.

Having specified the Lagrangian, we need to say something about the metric that these
modified theories might describe. As in the case of unmodified Plebanski theory, one may
declare the two-forms $B^i$ to be self-dual, which fixes the conformal class of the
metric to be the Urbantke one (\ref{urb}). However, now there exist many different
choices for the volume form. The most natural choice seems to be the one suggested by the
pure-connection formulation, which is to be described below.

Lagrangian (\ref{pleb-modified}), with all fields taken complex-valued, describes
modified complexified GR\@. For Riemannian and split signatures, one makes the same
respective choice of reality conditions as in the unmodified Plebanski theory. Thus, in
these signatures, the modified gravity theories under investigation have an honest
existence. For Lorentzian signature, one might hope that choice (\ref{reality-B}) remains
to be compatible with the dynamics of the theory after its modification. However, this is
known to be not true in many situations. Thus, no general reality conditions are known at
present that would allow for a physical interpretation of the modified theories
(\ref{pleb-modified}). However, in some special situations, conditions (\ref{reality-B}),
requiring the reality of the metric, are compatible with the dynamics. In these cases,
one does obtain real metrics with Lorentzian signature as solutions to modified theories.
These solutions exhibit interesting properties, and this is the reason why it looks
justified to study them. Whether solutions of this kind can be imbedded into some future
physical theory, to be defined by an appropriate generic choice of the reality
conditions, is still an open problem.

The family of theories (\ref{pleb-modified}) was previously studied in
\cite{Krasnov:2007uu, Krasnov:2007ky, Krasnov:2008sb, Krasnov:2010tt}. Due to a large
number of independent field components present in (\ref{pleb-modified}), calculations
with this theory are quite cumbersome. The pure-connection formulation introduced in
\cite{Krasnov:2011up} and \cite{Krasnov:2011pp} simplifies equations by eliminating most
of the field components. It thus simplifies the study of these modified gravity theories
considerably. The present paper is the first one where the effects of the modified
gravity are analysed directly in the simpler pure-connection formulation.

\subsection{An alternative parametrisation}

As a step towards the pure-connection formulation, we now describe an equivalent, but
slightly different parametrisation of the modified theories (\ref{pleb-modified}).
Lagrangian (\ref{pleb-modified}) can be written as
\beq\label{pleb-M}
S[B,A,M] = \im \int B^i \wedge F^i - \frac{1}{2} M^{ij} B^i \wedge B^j \, ,
\eeq
where the symmetric nondegenerate matrix $M^{ij}$ is subject to the constraint
\beq\label{M-constraint}
\tr M = \tilde \Lambda (M) \, .
\eeq
Here, $\tilde \Lambda (M)$ is some ${\rm SO}(3)$-invariant function of the matrix $M$. It
can be assumed without loss of generality that $\tilde \Lambda (M)$ only depends on the
two invariants $\tr (M^2)$ and $\tr (M^3)$, because if it also depended on $\tr M$, then
the constraint equation (\ref{M-constraint}) written in the form
\beq \label{M-conn}
\tr M = \tilde \Lambda \left( \tr M, \tr (M^2), \tr (M^3) \right)
\eeq
could be solved with respect to $\tr M$, giving birth to a new equation of the form
(\ref{M-conn}), in which a new function $\tilde \Lambda (M)$ would be ($\tr
M$)-independent.

To make a link to the previous parametrisation, we introduce a new matrix variable $\Psi$
via the relation
\beq\label{M-Psi}
\Psi^{ij} = M^{ij} - \frac{1}{3} \tilde \Lambda (M) \delta^{ij}\, .
\eeq
This turns constraint (\ref{M-constraint}) into the relation $\tr \Psi = 0$, and
establishes equivalence between theories (\ref{pleb-M}), (\ref{M-constraint})  and
(\ref{pleb-modified}) with $\Lambda (\Psi (M)) = \tilde \Lambda (M)$. Because of this
equivalence, we can identify $\Lambda (\Psi)$ and $\tilde \Lambda (M)$ and remove the
tilde from the latter to simplify the notation.

\subsection{The pure-connection formulation}

The pure-connection formalism for GR, as well as for the modified theories
(\ref{pleb-modified}), was developed in \cite{Krasnov:2011up, Krasnov:2011pp}. Here, it
will be instructive and useful to derive it from the first-order description
(\ref{pleb-M}), (\ref{M-constraint}), which is equivalent to (\ref{pleb-modified}).

The pure-connection formulation of (\ref{pleb-modified}), or, alternatively, of
(\ref{pleb-M}), is obtained by integrating out all the fields apart from the connection.
As the first step, one solves for the two-form fields:
\beq\label{B-F}
B = M^{-1} F\, ,
\eeq
where $M^{-1}$ is the matrix inverse of $M^{ij}$. Substituting this back into
(\ref{pleb-M}), we obtain
\beq
S[A,M] = \frac{\im}{2} \int \tr  \left( M^{-1} F\wedge F\right) \, .
\eeq
By choosing an arbitrary volume form  ${\mathcal V}$, we define a matrix $X^{ij}$ of
scalars related to the curvature wedge products as follows:
\beq\label{X}
F^i \wedge F^j = X^{ij} {\mathcal V} \, .
\eeq
The action then becomes
\beq\label{S-AM}
S[A,M] = \frac{\im}{2} \int \tr \left( M^{-1} X\right) {\mathcal V} \, .
\eeq

Now we can integrate out the matrix $M$ by solving its (algebraic) field equation
obtained by varying this action subject to constraint (\ref{M-constraint}). This leads to
the pure-connection action
\beq\label{S-A-M}
S[A] = \frac{\im}{2} \int \tr \left( M^{-1}(X) X\right) {\mathcal V} \, .
\eeq
This can alternatively be written as
\beq\label{S-pure-conn}
S[A] = \frac{\im}{2} \int f(X) {\mathcal V} \, ,
\eeq
which is the usual form of the pure-connection formulation as it is described, e.g., in
\cite{Krasnov:2011up}. From these expressions, we can identify the function $f(X)$ of the
pure-connection formulation as
\beq\label{f-X}
f(X) = \tr (M^{-1}(X) X) \, .
\eeq
Since action (\ref{S-A-M}) is stationary with respect to variations of $M$ [this gives
the equation used to determine $M^{-1}(X)$], we also see that
\beq\label{M-1}
\frac{\partial f}{\partial X} = M^{-1}(X) \, .
\eeq
Thus, the only remaining unsolved field equation of (\ref{pleb-M}),
\beq
d_A \left( M^{-1}(X) F\right) = 0 \, ,
\eeq
becomes the pure-connection formulation field equation
\beq\label{eqn-pure-conn}
d_A \left( \frac{\partial f}{\partial X} F\right) = 0 \, .
\eeq
This is a second-order partial differential equation for the connection.

Note that the SO(3)-invariant function $f (X)$ defined in (\ref{f-X}) has an important
property of being a homogeneous function of its argument: $f (z X) = z f (X)$ for any
complex number $z$.  This is because $X$ enters the action (\ref{S-AM}) linearly, so that
the solution $M^{-1} (X)$ in (\ref{f-X}) is invariant with respect to a rescaling of $X$.
In view of this property, and using definition (\ref{X}), we can express the Lagrangian
in (\ref{S-pure-conn}) symbolically as
\beq
f(X) {\mathcal V} = f \left( F \wedge F \right) \, .
\eeq

General Relativity (with a non-zero cosmological constant $\Lambda_0$) corresponds to a
particular choice of the function $f (X)$ in (\ref{S-pure-conn}); see
\cite{Krasnov:2011pp}:
\beq \label{GR}
f_{\rm GR} (X) = \frac{1}{\Lambda_0} \left( \tr \sqrt{X} \right)^2 \, .
\eeq
Another choice of a homogeneous SO(3)-invariant function $f (X)$ gives modification of
General Relativity. Such modifications can be regarded as GR with arbitrary function of
the curvature $F$ added to the Lagrangian, as discussed above.

\subsection{The metric from the connection}

To clarify the statement that Lagrangian (\ref{GR}) describes GR, we should specify the
metric $g$ that solves Einstein equations when $A$ satisfies its Euler-Lagrange
equations. This can be done directly in the pure-connection language, without referring
to the original Plebanski construction involving the two-form fields $B^i$.

As in the first-order Plebanski formalism, the metric is constructed in two steps. First,
we notice that, when the field equations (\ref{B-F}) are satisfied, the conformal class
of metrics in which $B^i$ are self-dual is the same as the one in which $F^i$ are
self-dual, since both sets of two-forms span the same three-dimensional subspace in the
space of all two-forms. Thus, instead of using (\ref{urb}), we can obtain the conformal
class of the metric directly from the curvature of the connection:
\beq\label{urb-curv}
g_{\alpha\beta} \sqrt{\det g}  \sim \tilde \epsilon^{\mu\nu\rho\sigma} \epsilon_{ijk}
F^i_{\mu\nu} F^j_{\rho\alpha} F^k_{\sigma\beta} \, .
\eeq
For complex-valued connection components $A^i$, the conformal class obtained is, in
general, complex (i.e., does not contain any real-valued metric). The reality conditions
that ensure that the conformal class is that of a real Lorentzian metric can be stated as
\beq\label{reality}
F^i \wedge \bar F^j = 0 \, .
\eeq
All this parallels the above discussion; we have only replaced the two-forms $B^i$ by
$F^i$.

In the second step, one specifies a particular metric in the conformal class already
defined. To do so, it suffices to fix the associated volume form. The volume form ${\cal
V}_A$ that gives the metric satisfying the Einstein equations in the case of theory
(\ref{GR}) is determined by
\beq\label{volume}
-2 \im \Lambda_0^2 {\cal V}_A = \Lambda_0 f_{\rm GR}(F\wedge F) \, .
\eeq
The factor of the imaginary unit is needed in this relation for the same reason as in
(\ref{S-pure-conn}). Thus, in the case of pure-connection description of GR, the action
of the theory (\ref{S-pure-conn}) is just the total volume of the space, which is very
appealing.

For a modified gravity theory with a general function $f \left( F\wedge F \right)$ as a
Lagrangian, we do not yet know what the correct `physical' metric would be. However, by
analogy with GR, we could assume that the action is again the total volume calculated
with respect to this metric. This would imply that the metric volume form is a multiple
of $f \left( F\wedge F \right)$. It is this metric that we are going to study below.

\subsection{Mixed parametrisation}

It is not hard to pass from the Plebanski formulation of GR with $\Lambda\not=0$ to the
pure-connection formulation of GR and to obtain the Lagrangian function (\ref{GR}). It
turns out to be surprisingly hard, however, to characterise modifications even of a
simple type (\ref{alpha-family}) in the pure-connection language. It is also clear from
(\ref{GR}) that the pure-connection description involves the operation of taking the
square root of a matrix, which requires specifying its branch. Another issue with the
pure-connection formulation is that it only makes sense for $\Lambda\not =0$.

To overcome these difficulties, in this subsection we introduce a new, `mixed'
parametrisation that lies half-way between the Plebanski first-order description
(\ref{pleb-M}) and the pure-connection description (\ref{S-pure-conn}) and that combines
the advantages of both. The idea is to avoid solving the field equations for $M$, and
work with both $M$ and $A$ as independent variables.

Let us see how this can work in practice. As a first step, we replace the constrained
variational problem for $M$ in (\ref{S-AM}) by an unconstrained one, with an additional
Lagrange multiplier $\mu$ imposing the constraint:
\beq\label{S-AM-mu}
S[A,M,\mu] = \frac{\im}{2} \int \left[ \tr \left( M^{-1} X\right) + \mu \left( \tr (M) -
\Lambda(M)\right) \right] {\mathcal V} \, .
\eeq
In the context of unmodified GR, the idea of using a Lagrange multiplier to impose the
tracelessness constraint, to our knowledge, was first implemented in \cite{Beke}.
Variation with respect to $M$ then gives the equation
\beq\label{MX-eqn}
M^{-1} X = \mu \left( M - \frac{\partial \Lambda(M)}{\partial M} M\right) \, .
\eeq
To get to the pure-connection formulation, we are supposed to solve this, and the
constraint equation (\ref{M-constraint}), for $M$ and $\mu$ in terms of $X$. However,
this is very difficult to do in general. Only in the case of GR, where $\Lambda$ does not
depend on $M$, one can do this and immediately obtain (\ref{GR}). For this reason, let us
instead solve equation (\ref{MX-eqn}) with respect to $X$. Taking the trace of this
equation, and using the constraint on $M$, as well as definition (\ref{f-X}), we can
eliminate the Lagrange multiplier $\mu$ and get
\beq\label{X-M}
X = f(X) \frac{ M^2 - M \frac{\partial \Lambda(M)}{\partial M} M}{ \Lambda(M) - \tr
\left[ \frac{\partial \Lambda(M)}{\partial M} M\right] } \, .
\eeq
Thus, the matrix $X$ is determined by $M$ up to scale. Then, if we interpret the field
equation
\beq\label{M-pde}
d_A (M^{-1} F)=0
\eeq
as a partial differential equation for $M$ and solve it, we will find $X$ from
(\ref{X-M}). This is the strategy we will follow below to solve the field equations in
the case of homogeneous anisotropic cosmology.

Let us also give the form of relation (\ref{X-M}) in the case of parametrisation by using
variable (\ref{M-Psi}) and function $\Lambda (\Psi)$ in (\ref{pleb-modified}). We have
\beq
X = f(X) \frac{\left( \Psi + \frac{1}{3} \Lambda(\Psi) {\rm Id}\right)^2 \left( {\rm Id}
- \frac{\partial\Lambda}{\partial \Psi}\right)}{\Lambda(\Psi) - \tr  \left(
\frac{\partial\Lambda}{\partial \Psi} \Psi\right)} \, ,
\eeq
where Id denotes the identity matrix. In obtaining the above formula we have used the
fact that the matrices $\Psi$ and $\partial \Lambda/\partial \Psi$ commute.

The case $\Lambda(\Psi) \equiv 0$, which corresponds to GR without the cosmological
constant, needs to be considered separately. This theory does not have pure-connection
formulation, but one can still solve the field equations in the mixed parametrisation
proposed here. In this case, equation (\ref{MX-eqn}) gives
\beq\label{X-M-zero}
\Psi^{-1} X = \frac{1}{3} \Psi\, \tr (\Psi^{-2}X) \, .
\eeq
In particular, this equation implies that the function $f(X)=\tr (\Psi^{-1}(X) X) \equiv
0$ in the case of GR without the cosmological constant. We analyse the Bianchi~I
cosmology for this case in the Appendix.

In either of the two cases, one needs to solve the coupled system of equations
(\ref{X-M}), (\ref{M-pde}) or (\ref{X-M-zero}), (\ref{M-pde}) (with $M = \Psi$) and find
the components of the connection. So, this is still a connection formulation with the
connection playing the role of the main field. However, there are now additional
auxiliary fields $\Psi$. One of the benefits of this mixed formulation is that, even in
the case of GR, there is no need to take square roots. And, as we shall see below, the
coupled system of equations can be solved, at least in special situations.

\subsection{Gravitational Instantons}

Before we proceed to the analysis of the homogeneous anisotropic field configurations as
described by our modified gravity theories, we make one further remark about the
gravitational instantons in our formalism. This subsection can be skipped on the first
reading.

Let us rewrite the mixed parametrisation (\ref{S-AM-mu}) of our theories as
\beq \label{g-M}
S[A,M,\mu] = \frac{\im}{2} \int \left[ \tr \left( M^{-1} X\right) + \mu \,g(M)
\right]{\mathcal V}\, ,
\eeq
where we have introduced a scalar-valued function of matrix $M$\,:
\beq
g(M) := \tr M - \Lambda (M) \, .
\eeq
The role of the Lagrange multiplier $\mu$ is to impose a single constraint on the matrix
$M^{ij}$. This constraint can be viewed as defining a codimension-one surface $g(M)=0$ in
the space of matrices $M$. By the Hamiltonian analysis of this class of theories, this
constraint can be shown to be, in fact, the Hamiltonian constraint in disguise. As we
have seen above, the function
\beq\label{GR-l}
g_{\rm GR}(M) = \tr M - \Lambda_0
\eeq
with constant $\Lambda_0$ gives the description of GR with the cosmological constant
$\Lambda_0$.

We now note that, $M$ being an auxiliary field, any field redefinition of $M$ in
(\ref{g-M}) should give an equivalent description of the theory. Thus, in particular, the
Lagrangian
\beq\label{S-AtM-mu}
S[A,\tilde{M},\mu] = \frac{\im}{2} \int \left[ \tr \left( \tilde{M} X \right) + \mu
\,\tilde{g} (\tilde{M}) \right]{\mathcal V} \, ,
\eeq
where $\tilde{M}=M^{-1}$, and $\tilde{g} (\tilde M) = g (\tilde M^{-1})$ should provide
an equivalent description.

It is then interesting to note that, by taking
\beq
\tilde{g}_{\mbox{\scriptsize self-dual}}(\tilde{M}) = \tr \tilde{M} - \Lambda_0
\eeq
with constant $\Lambda_0$, one obtains a Lagrangian describing self-dual gravity. Indeed,
varying (\ref{S-AtM-mu}) with respect to $\tilde{M}$, we obtain the equations
\beq
X^{ij}\sim \delta^{ij} \, ,
\eeq
which are known to provide the correct description of the gravitational instantons in the
pure-connection formalism.  In terms of the original action (\ref{g-M}), this means
choosing the function $g (M)$ in the form
\beq\label{inst-l}
g_{\mbox{\scriptsize self-dual}} (M) = \tr \left( M^{-1} \right) - \Lambda_0 \, .
\eeq

Thus, our mixed parametrisation is flexible enough to incorporate GR, modified gravity
theories, and self-dual gravity. One just constraints the auxiliary matrix field onto
different codimension-one surfaces. It is interesting that GR (\ref{GR-l}) and the
instanton sector (\ref{inst-l}) go one into another by the replacement $M \to M^{-1}$ in
the constraint equation.

\section{Bianchi~I connections}
\label{sec:Bianchi}

We now start exploring how the familiar general-relativistic Kasner solution arises in
the above pure-connection formulation, and what the modifications of gravity entail.

The following ansatz for the Bianchi~I connection is motivated in the Appendix:
\beq
A^i = \im \g_i(\tau) dx^i \, .
\eeq
Note that there is no summation over $i$ on the right-hand-side of this equation. Here,
$\g_i(\tau)$ are three functions of an arbitrary time coordinate $\tau$, while $x^i$ are
the Cartesian coordinates on the spatial slices (surfaces of homogeneity). The
corresponding curvature two-form is
\beq \label{Fi}
F^i = d A^i + \frac12 \epsilon^{ijk} A_j \wedge A_k = \ri \dot \g_i d \tau \wedge d x^i -
\frac12 \epsilon^{ijk} \g_j \g_k d x^j \wedge d x^k \, ,
\eeq
where an overdot denotes derivative with respect to $\tau$. Calculating the wedge
product, we obtain
\beq \label{Xi}
F^i \wedge F^j = - 2 \ri \delta^{ij} X_i \g \cV_c \, ,
\eeq
where $\cV_c = d\tau \wedge dx^1 \wedge dx^2 \wedge dx^3 $ is the coordinate volume form,
$\g =\g_1 \g_2 \g_3$, and
\beq \label{eq-g}
X_i = \frac{\dot \g_i}{\g_i} \, .
\eeq
If we select the volume form $\cV$ in (\ref{X}) to be
\beq
\cV = - 2 \ri  \g \cV_c \, ,
\eeq
then $X^{ij} = {\rm diag}\, \left(X_1, X_2, X_3 \right)$.

\subsection{Evolution equations in the pure-connection parametrisation}

The reason for the above choice of the volume form defining the matrix $X^{ij}$ is that
the pure-connection formulation equation (\ref{eqn-pure-conn}) reduces to the system
\beq\label{eq-diag}
\left( \frac{\partial f}{\partial X_i}\right)^\cdot  = f (X) -  \frac{\partial
f}{\partial X_i} \sum_j X_j \, ,
\eeq
which is a system of first-order differential equations for $X_i$. A derivation is given
in the Appendix (in more generality), but, for the present situation with diagonal matrix
$X$, it is quite easy to obtain these equations directly. Specialising to the case of the
function $f (X)$ given by (\ref{GR}), it is not hard to obtain the familiar GR solution;
see Appendix.

An alternative form of equations (\ref{eq-diag}) is obtained by multiplying these
equations by $\g =\g_1 \g_2 \g_3$ and using definition (\ref{eq-g}).  Equations
(\ref{eq-diag}) then reduce to
\beq \label{eq-new}
\left( \frac{\partial f}{\partial X_i} \g \right)^\cdot = f(X) \g \, .
\eeq
This form will be very convenient for analysing the case of arbitrary $f(X)$.

\subsection{Evolution equations in the parametrisation with auxiliary fields}

Let us also give the form of the evolution equations arising in the mixed
parametrisation, which involves both the connection and the matrix $M$. The equation in
question is then (\ref{M-pde}). Given that the matrix $M$ is the inverse of the matrix of
first derivatives of the function $f(X)$ [see (\ref{M-1})], it is diagonal whenever $X$
is diagonal: $M^{ij}={\rm diag}(M_1,M_2,M_3)$. Using relation (\ref{f-X}), one can write
equations (\ref{M-pde}) in a form similar to (\ref{eq-new}):
\beq\label{eq-M}
\left( M_i^{-1} \g \right)^{\cdot} = \tr \left( M^{-1} X \right) \g \, .
\eeq
This form will be most convenient for the analysis of modifications in the
parametrisation by a function $\Lambda(\Psi)$.

\subsection{The metric}

Before we begin our analysis of the evolution equations given above, it is useful to
compute the metric determined by the connection. To this end, one can follow the
previously described procedure by first computing the Urbantke metric (\ref{urb-curv})
and then conformally rescaling it so that the volume form becomes a multiple of $f \left(
F\wedge F \right)$. A simpler method is to directly look for a metric that makes the
curvature forms (\ref{Fi}) self-dual.

We are looking for the metric in the Bianchi~I form
\beq \label{metric}
ds^2 = - N^2 (\tau) d\tau^2 + \sum_i a_i^2 (\tau) \left( dx^i\right)^2 \, .
\eeq
Calculating the dual $^*F^i$ of the curvature two-forms (\ref{Fi}) with respect to this
metric, we have
\beq
^*F^i = - \ri \dot \g_i \frac{N a_1 a_2 a_3}{2 N^2 a_i^2} \epsilon_{ijk} dx^j \wedge dx^k
- \frac12 \epsilon^{ijk} \g_j \g_k \frac{N a_1 a_2 a_3}{a_j^2 a_k^2} \epsilon_{jkl} d\tau
\wedge dx^l \, .
\eeq
The requirement $^*F^i = {} \pm \ri F^i$ gives the condition
\beq
{} \pm \dot \g_1 = \frac{N a_1}{a_2 a_3} \g_2 \g_3 \, , \quad \mbox{etc} \, ,
\eeq
from which we get
\beq \label{aN}
\frac{a_1^2}{N^2} = \frac{\g_1^2}{X_2 X_3} \, , \quad \mbox{etc}\, .
\eeq

Another equation for determining the metric is obtained by fixing the metric volume form
\beq \label{volmet}
\cV_m = N a_1 a_2 a_3 d\tau \wedge dx^1 \wedge dx^2 \wedge dx^3 = N a_1 a_2 a_3 \cV_c\, .
\eeq
By our prescription, this should be equal to a multiple of $f \left( F\wedge F \right)$:
\beq
-2\im \Lambda_0 \cV_m = f \left( F\wedge F \right) \, ,
\eeq
where $\Lambda_0$ is some parameter of dimension $1/L^2$, later to be identified with the
cosmological constant. Using (\ref{Xi}), we get
\beq\label{volume-metric-f}
\Lambda_0 N a_1 a_2 a_3 = f(X) \g \, .
\eeq
Combining this equation with (\ref{aN}), we have
\beq \label{lapse}
N^2 =  \left( \frac{f^2(X) \prod_i X_i^2}{\Lambda_0^2} \right)^{1/4}\, , \qquad a_1^2 =
N^2 \frac{\g_1^2}{X_2 X_3}  \, , \quad \mbox{etc} \, ,
\eeq
where the first relation is written in the form indicating that there are, in general,
four possible branches, two of them imaginary. If we require that the metric be real, and
that the signature of the $\tau$ coordinate be negative, then the final expression for
the metric is
\beq \label{canonmet}
ds^2 =  \sqrt{ \left| \frac{f(X) \prod_i X_i}{\Lambda_0 } \right| } \left[ - d \tau^2 +
\prod_j X_j^{-1}  \sum_k \g_k^2 X_k \left(d x^k \right)^2 \right] \, .
\eeq
Note that this metric is time-reparametrisation invariant, as it should be.

\section{Solution in the general case}
\label{sec:general}

One of the miracles of the pure-connection formulation of gravity under consideration is
that it allows one to write the {\em general\/} solution to the problem at hand for an
arbitrary theory, i.e., for an arbitrary choice of the function $f (X)$. This becomes
possible by using a clever choice of the time variable.

We begin with solution for the case of general $f(X)$ and then specialise to GR and to a
particular modified gravity theory. Solution of GR in which one works in the physical
time from the beginning is also possible. For completeness, it is presented in the
Appendix.

In the previous section, we arrived at evolution equations in the form (\ref{eq-new}). By
using time-reparametrisation freedom, it is always possible to choose the time variable
$\tau$ in such a way that
\beq \label{tcon}
f \g = {\rm const} \, .
\eeq
The geometric significance of this choice is that this is the time coordinate in which
the metric volume form is proportional to the coordinate volume form, i.e., $\sqrt{\det
g} = Na_1 a_2 a_3 = {\rm const}$. This is clear from (\ref{volume-metric-f}).

With this choice, equation (\ref{eq-new}) can be integrated to give an implicit solution
for $X (\tau)$:
\beq \label{Xsol}
\frac{\partial f (X)}{\partial X_i} = f (X) \left( \tau - \tau_i \right) \, ,
\eeq
where $\tau_i$ are arbitrary integration constants.  The homogeneity of the function $f
(X)$ implies another relation
\beq \label{Xtime}
\sum_i X_i \left( \tau - \tau_i \right) = 1 \, .
\eeq
Equations (\ref{Xsol}) and (\ref{eq-g}) give a complete solution to the problem for an
arbitrary theory from our class. We now give some general analysis of the solution
obtained, and then consider some specific functions $f (X)$.

\subsection{De Sitter solution}

Consider $\tau \to \infty$, and assume that $f(X)\tau$ remains constant as $\tau\to \infty$. Then
equation (\ref{Xsol}) implies that all derivatives $\partial f (X)
/ \partial X_i$ become mutually equal.  The symmetry of the function $f (X)$, in turn,
implies that all $X_i$ become equal to each other in this limit.  Relation (\ref{Xtime})
then gives the solution
\beq
X_i \approx \frac{1}{3 \tau} \quad \mbox{as \ $\tau \to \infty$} \, .
\eeq
The homogeneity of $f(X)$ then justifies the assumption $f(X)\tau \to {\rm const}$ that
we made in deriving this solution.

The corresponding metric describes the De~Sitter spacetime. Indeed, we have
$f(X)=f_0/\tau$, where $f_0$ is a constant. Then, by rescaling the spatial coordinates,
we can always choose the solution in the form $\g_i = \tau^{1/3}$. Then metric
(\ref{canonmet}) becomes
\beq
ds^2 = \sqrt{\frac{3 f_0}{\Lambda_0}} \left( - \frac{d \tau^2}{9 \tau^2} + \tau^{2/3}
dr^2 \right) = \sqrt{\frac{3 f_0}{\Lambda_0}} \left( - d t^2 + \re^{2 t} dr^2 \right) \,
,
\eeq
where $\tau = \re^{3 t}$ is the time coordinate change, and $d r^2 = \sum_i \left( d x^i
\right)^2$. This is nothing but the De~Sitter metric, which is thus the solution of any
modified theory.

\subsection{Integration constants}

Without loss of generality, one can shift the time variable so that
\beq \label{tini}
\sum_i \tau_i = 0 \, .
\eeq
Second, apart from the trivial case $\tau_i = 0$ for all $i$, which gives the De~Sitter
solution, by the remaining freedom of time rescaling, which does not violate
(\ref{tcon}), we can achieve the condition
\beq\label{gf-2}
\sum_i \tau_i^2 = 2 \, .
\eeq
This normalization is convenient because squaring (\ref{tini}) we can rewrite
(\ref{gf-2}) as
\beq
\tau_1 \tau_2 + \tau_2 \tau_3 + \tau_3 \tau_1 = -1.
\eeq

Without loss of generality, we can arrange the integration constants so that
\beq\label{t-order}
\tau_3 \leq \tau_2 \leq \tau_1 \, .
\eeq
Because of condition (\ref{tini}), we have $\tau_3 < 0 < \tau_1$. When $\tau_2 = \tau_1$,
we have $\tau_2 = \tau_1 = 1/\sqrt{3}$ and $\tau_3=-2/\sqrt{3}$. This is the largest
absolute value that $\tau_3$ can reach. In the opposite extreme $\tau_2=\tau_3$ we have
$\tau_2=\tau_3=-1/\sqrt{3}$ and $\tau_1=2/\sqrt{3}$, which is the largest value $\tau_1$
can reach. All in all, we have
\beq
\tau_c \leq \tau_1 \leq 2 \tau_c\, , \qquad -\tau_c \leq \tau_2 \leq \tau_c\, , \qquad
-2\tau_c \leq \tau_3 \leq -\tau_c\, ,
\eeq
where $\tau_c=1/\sqrt{3}$.

\subsection{Solution in parametrisation with auxiliary matrix $M$}

From now on, we will work in the mixed parametrisation with auxiliary matrix $M$. The
evolution equation in this case takes the form (\ref{eq-M}). Choosing the same time
variable as above, namely, the one that satisfies (\ref{tcon}), we immediately obtain the
solution
\beq
M^{-1}  = f (X) T \, ,
\eeq
where $T^{ij} = \delta^{ij} (\tau - \tau_i)$ is another diagonal matrix.  This, in fact,
is the same solution as (\ref{Xsol}); we simply refrained from solving the relation
between $M$ and $X$. From this, we have
\beq \label{M}
\tr M = \Lambda (M) = \frac{\tr (T^{-1}) }{f (X)} \, , \qquad M = \Lambda (M)
\frac{T^{-1}}{\tr (T^{-1})} \, .
\eeq
Equation (\ref{X-M}) then becomes
\beq \label{Xfin}
X  = \frac{ T^{-2} \left( {\rm Id}  - \frac{\partial \Lambda (M)}{\partial M}\right)}{\tr
\left[ T^{-1} \left( {\rm Id}  - \frac{\partial \Lambda (M)}{\partial M} \right) \right]
} \, .
\eeq
The second equation in (\ref{M}) enables one to find $M$ as a function of time.
Substituting it into (\ref{Xfin}), one finds the solution for $X$.

\subsection{Solution in parametrisation with auxiliary matrix $\Psi$}

By expressing the partial derivatives in (\ref{Xfin}) in terms of $\Psi$ introduced in
(\ref{M-Psi}), one easily obtains a solution in the form
\beq \label{Xp}
X  = \frac{ T^{-2} \left(  {\rm Id} - \frac{\partial \Lambda (\Psi)}{\partial
\Psi}\right)}{\tr \left[ T^{-1}\left( {\rm Id} - \frac{\partial \Lambda (\Psi)}{\partial
\Psi} \right) \right]} \, .
\eeq
In the $\Psi$ parametrisation, the second equation in (\ref{M}) is written as
\beq \label{P}
\Psi = \Lambda (\Psi) \left[ \frac{T^{-1}}{\tr (T^{-1})} -  \frac{1}{3}{\rm Id} \right]
\, .
\eeq
It is to be solved with respect to $\Psi$, with the result to be substituted into
(\ref{Xp}).

\section{The case of GR}
\label{sec:GR}

\begin{figure}
\centering
\includegraphics[width=\textwidth]{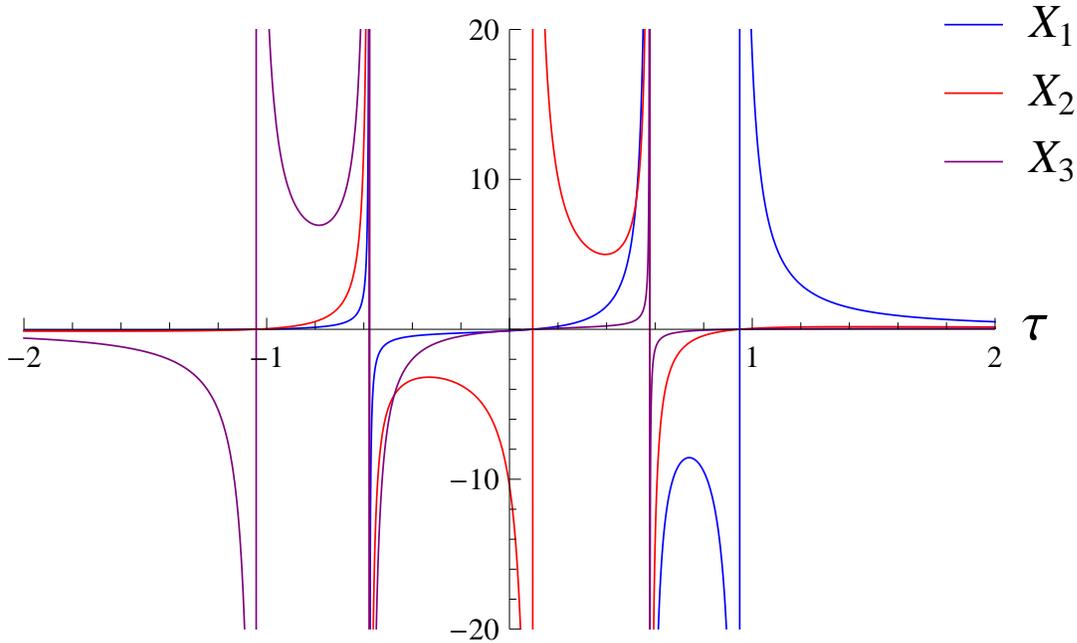}
\caption{Behaviour of the variables $X_i$ given by (\ref{X-GR}) in General Relativity. Each
variable $X_i$ has a pole at respective $\tau = \tau_i$.  These are just special points
of the solution, with all connection and metric components remaining finite. Furthermore,
all variables $X_i$ have common poles at $\tau = \pm \tau_c$. These are the singular
points for both the connection and the metric. The region `between' two Kasner
singularities describes a solution with negative $\Lambda$, while the two outer regions
describe solutions with $\Lambda>0$.} \label{fig:GR}
\end{figure}

Here, we obtain the solution of GR in this time variable. In the case $\Lambda(\Psi)
\equiv \Lambda_0 = {\rm const}$, one gets
\beq\label{X-GR}
X_i^{\rm GR} = \frac{1}{s_1 (\tau-\tau_i)^2}\, , \quad {\rm or} \quad X_1^{\rm GR} =
\frac{(\tau-\tau_2)(\tau-\tau_3)}{(3\tau^2-1)(\tau-\tau_1)}\, , \quad {\rm etc} \, ,
\eeq
where
\beq\label{s1}
s_1 \equiv \tr (T^{-1}) \equiv \sum_i \frac{1}{\tau-\tau_i} = \frac{3\tau^2-1}{\prod_i
(\tau-\tau_i)} \, ,
\eeq
and we have used both (\ref{tini}) and (\ref{gf-2}). The quantities $X_i^{\rm GR}$ have
simple poles at $\tau = \tau_i$, and all blow up as $\tau\to \pm 1/\sqrt{3}$, which
corresponds to the Kasner singularity. This behaviour is illustrated in
Fig.~\ref{fig:GR}.

Let us also write the corresponding metric components; see (\ref{lapse}). According to
(\ref{M}), we have
\beq\label{f-x}
f(X) = \frac{s_1}{\Lambda_0}  \, .
\eeq
Thus,
\beq
\frac{f(X) X_1 X_2 X_3}{\Lambda_0} = \frac{1}{\Lambda_0^2 (3\tau^2-1)^2}\, , \qquad
\frac{f(X) X_1}{\Lambda_0 X_2 X_3} = \frac{(3\tau^2-1)^2}{\Lambda_0^2 (\tau-\tau_1)^4}\,
,
\eeq
and similarly for the other components. All expressions are manifestly positive, so
taking the square root, we have
\beq\label{lapse-gr}
N^2 = \frac{1}{\Lambda_0(3\tau^2-1)} \, , \qquad a_i^2 = \g_i^2
\frac{3\tau^2-1}{\Lambda_0(\tau-\tau_1)^2} \, .
\eeq

In the time interval $\tau\in (-\tau_c,\tau_c)$, $\tau_c = 1 / \sqrt{3}$, instead of
taking the modulus of expressions to get non-negative metric components, we reverse the
sign of the cosmological constant $\Lambda_0$.  This is the correct interpretation, as
this time interval corresponds to a solution of GR with negative cosmological constant.

We now study this solution in more detail, and, in particular, integrate the equations
for $\g_i$ near the singularity.

\subsection{Behaviour near the poles}

When all three integration constants are different, the function $s_1(\tau)$ has three
simple poles at $\tau=\tau_i$, and two simple zeros at $\tau=\pm \tau_c$. Let us analyse
the behavior near the poles.

Consider, for example, the limit $\tau \to \tau_1$. In this case, we have $s_1 \sim
1/(\tau - \tau_1) \to \infty$. Solution (\ref{X-GR}) behaves as
\beq \label{Xs}
X_1 \approx \frac{1}{\tau-\tau_1}\, , \qquad X_2 \sim X_3 \sim \tau - \tau_1 \, .
\eeq
This is an integrable behavior, with $\g_1 \to 0$ and $\g_2$ and $\g_3$ finite as $\tau
\to \tau_1$. We thus see that all $X_i$ change sign at $\tau=\tau_1$.

Let us determine the behaviour of the components (\ref{lapse}) of the canonical metric
(\ref{canonmet}) at this point.  Integrating the first equation in (\ref{Xs}), we obtain
\beq
\g_1 \sim \tau - \tau_1 \, ,
\eeq
while $\g_2$ and $\g_3$ tend to constants.  So, the significance of the point
$\tau=\tau_1$ is in the fact that one of the connection components $\g_1$ passes through
zero there.

Now, using this behaviour we see that the metric lapse function as well as the scale
factors (\ref{lapse}) are finite and regular as $\tau \to \tau_1$. So, the $\tau=\tau_1$
is just a special point in the solution.

\subsection{Behaviour near the singularity}

At the singularity $\tau = \tau_c=1/\sqrt{3}$, the function $s_1$ has a simple zero.
Thus, we have
\beq
X_{1}\approx -
\frac{(\tau_c-\tau_2)(\tau_c-\tau_3)}{2\sqrt{3}(\tau_1-\tau_c)(\tau-\tau_c)} \, , \quad
\mbox{etc} \, , \qquad f(X) \sim \tau-\tau_c \, .
\eeq
Integrating (\ref{eq-g}), we get
\beq \label{gsin}
\g_1 \sim (\tau-\tau_c)^{-
\frac{(\tau_c-\tau_2)(\tau_c-\tau_3)}{2\sqrt{3}(\tau_1-\tau_c)}} \, , \quad {\rm etc} \,
.
\eeq
We thus see that the lapse function (\ref{lapse-gr}) diverges, while the scale factors behave as
\beq\label{scale-gr}
a_i^2 \sim (\tau-\tau_c)^{p_i} \, ,
\eeq
where
\beq\label{ps}
p_1 = 1 - \frac{(\tau_c-\tau_2)(\tau_c-\tau_3)}{\sqrt{3}(\tau_1-\tau_c)} \, ,  \quad {\rm
etc} \, .
\eeq
These exponents satisfy
\beq\label{exponents}
p_1+p_2+p_3=1 \, , \qquad p_1 p_2+p_2 p_3+p_3 p_1=0 \, .
\eeq
From (\ref{lapse-gr}) we see that the physical time near the singularity is $t \sim
\sqrt{\tau-\tau_c}$, and thus the behaviour (\ref{scale-gr}) is the usual Kasner one,
\beq\label{Kasner-met}
a_i^2 \sim t^{2p_i} \, ,
\eeq
with the correct exponents (\ref{exponents}).

Note that the components of the gauge field (\ref{gsin}) all diverge at the singularity,
so this is a true singularity not only of the canonical metric (\ref{canonmet}) but also
of the fundamental gauge field.

\subsection{Summary of the GR solution}

We summarise the facts established above. As $\tau\to \infty$, we approach the De~Sitter
solution $X_i=1/{3\tau}$. As time decreases, at $\tau=\tau_1$ we encounter a special
point where $X_1$ has a simple pole, while $X_2$ and $X_3$ vanish. Below this point, all
$X_i$ change sign, as does $f(X)$. The component of the connection $\g_1$ vanishes at
this point, while the components $\g_2$ and $\g_3$ remain finite. All components of the
canonical metric (\ref{canonmet}) remain finite at this point.

As $\tau\to \tau_c$, we approach the Kasner singularity, with the functions $X_i$ all
negative near the singularity, and all having a simple pole there. The function $N^2$
also has a simple pole at this point. The scale factors $a_i^2$ exhibit the familiar
Kasner behaviour (\ref{scale-gr}).

We also note that the region $\tau > \tau_1$ is where the solution is guided by the
cosmological constant. Near the Kasner singularity, as $\tau \to \tau_c$, the Weyl
curvature becomes so strong that the cosmological constant does not play any role and can
be neglected.

Since the gauge field diverges at the singularity, the domain $- \tau_c < \tau < \tau_c$
in (\ref{X-GR}) can only be treated as another singular solution.  Let us first consider
the region $\tau_2 < \tau < \tau_c$. In order that the metric with components
(\ref{lapse-gr}) be of the usual signature there, we need to change the sign of the
cosmological constant $\Lambda_0$. Thus, the solution in this time interval is described
by GR with negative cosmological constant. The behaviour near $\tau_c$ is again Kasner.
The point $\tau_2$ again is a special point of the solution, in which $X_2$ has a simple
pole, while $X_1$ and $X_3$ have simple zeros. Thus, $\g_2$ passes through zero at this
point, with $\g_1$ and $\g_3$ remaining finite and nonzero. The metric components are all
finite and non-zero. As $\tau\to -\tau_c$, we encounter another Kasner singularity. Thus,
the $\Lambda_0<0$ part of the solution interpolates between two Kasner singularities.
There is no asymptotic anti-De~Sitter regime in this case.

For $\tau<-\tau_c$, we have another copy of asymptotically De~Sitter solution. We note
that all $X_i$ are positive near the singularity in this case, as is $f(X)$. There is a
Kasner singularity as $\tau \to -\tau_c$, and a special point at $\tau=\tau_3$ with
$\g_3$ vanishing and all $X_i$ and $f(X)$ changing sign. As $\tau\to -\infty$, we
approach another De~Sitter region. Since the time change $\tau \to - \tau$ makes the
region $\tau<-\tau_c$ mathematically equivalent to the asymptotically De~Sitter region
$\tau > \tau_c$ discussed above, it is clear that the Kasner exponents near the
singularity in the region $\tau < - \tau_c$ are obtained from (\ref{ps}) by the
replacement $\tau_i \to - \tau_i$.

\section{One-parameter family of modifications}
\label{sec:modified}

\subsection{Modified $\Lambda_0>0$ case}

Consider the one-parameter family (\ref{alpha-family}) with $\Lambda_0 > 0$ and small
modification $\alpha\Lambda_0\ll 1$. In this case, one expects that modification only
becomes significant in the region where the Weyl curvature is large, i.e., near the
singularity. This is correct, and, as we will see below, the singularity at $\tau=\pm
\tau_c$ gets resolved in the modified theory, leading to solutions in which the
fundamental connection field is regular all through the space-time.

To find solution for $X$, we need to solve equation (\ref{P}) for $\Psi (\tau)$ and
substitute the result into (\ref{Xp}).  Since, by virtue of (\ref{P}), $\Psi (\tau)$ is
determined by $\Lambda (\tau)$, it will be sufficient to find $\Lambda (\tau)$.  To find
$X$ from (\ref{Xp}) in theory (\ref{alpha-family}), we calculate
\beq\label{numerat}
{\rm Id} - \frac{\partial \Lambda}{\partial \Psi} = {\rm Id} +  \alpha \Psi = {\rm Id} +
\alpha \Lambda \left( \frac{ T^{-1}}{s_1}- \frac{1}{3} {\rm Id}   \right) \, ,
\eeq
\ber\label{den}
\tr \left[ T^{-1} \left( {\rm Id} - \frac{\partial \Lambda (\Psi)}{\partial \Psi} \right)
\right] = s_1 \left[ 1 + \alpha \Lambda \left( \frac{ s_2}{s_1^2} - \frac{1}{3} \right)
\right] \, ,
\eer
where, in addition to (\ref{s1}), we have introduced the function
\beq\label{s2}
s_2= \tr(T^{-2}) = \sum_i \frac{1}{(\tau - \tau_i)^2} = \frac{3\tau^4 + 6 \tau_1 \tau_2
\tau_3 \tau + 1}{\prod_i (\tau-\tau_i)^2} \, .
\eeq
Eventually, we have
\beq \label{Xp1}
X = \frac{(1 - \frac13 \alpha \Lambda) s_1 T^{-2} + \alpha \Lambda T^{-3}}{s_1^2 \left( 1
+ \alpha x \Lambda \right)} \, ,
\eeq
where
\beq \label{x}
x (\tau) \equiv \left( \frac{ s_2}{s_1^2} - \frac{1}{3} \right)= 2\frac{3\tau^2 + 9
\tau_1 \tau_2 \tau_3 \tau + 1}{3 \left( 3 \tau^2 - 1 \right)^2} \, .
\eeq

It remains to find $\Lambda (\tau)$.  Using (\ref{P}), we have
\beq
\Psi^2 = \Lambda^2 \left(  \frac{T^{-2}}{s_1^2} - \frac{2 T^{-1}}{3s_1} + \frac{1}{9}
{\rm Id} \right) \, , \qquad \tr (\Psi^2) =  \Lambda^2 \left( \frac{s_2}{s_1^2} -
\frac{1}{3} \right) \, ,
\eeq
and equation (\ref{alpha-family}) then produces the quadratic equation for $\Lambda
(\tau)$:
\beq \label{quadL}
-\frac{\alpha}{2} x (\tau) \Lambda^2 = \Lambda - \Lambda_0  \, .
\eeq

Consider first the region $|\tau| > \tau_c$. (The region $|\tau| < \tau_c$ is analysed in
Section \ref{sec:would-be}.) Note that the function $x(\tau)$, defined in (\ref{x}), is
nonnegative and approaches infinity as $|\tau| \to \tau_c$. Therefore, in order that the
quadratic equation (\ref{quadL}) always have a real solution for $\Lambda$, it is
necessary to demand that $\alpha > 0$. As $|\tau| \to \infty$, the function $x(\tau)$
tends to zero. In order that the solution for $\Lambda$ tend to the general-relativistic
value $\Lambda_0$ in this limit, one should take the positive root of the quadratic
equation (\ref{quadL}):
\beq \label{lambdax}
\Lambda = \Lambda_+ = \frac{1}{\alpha x} \left( \sqrt{1 + 2 \alpha x \Lambda_0}-1\right)
\, .
\eeq

Since $x$ and $\Lambda$ are both non-negative in the region $|\tau| > \tau_c$, the only
place where we possibly can encounter singularity in solution (\ref{Xp1}) is when $s_1$
turns to infinity or zero. The first option occurs as $\tau \to \tau_1$, at which point
the diagonal matrix $T^{-1}$ also becomes singular.  The second option occurs as $|\tau|
\to \tau_c = 1/\sqrt{3}$ (this is a singular point in GR)\@.  Let us consider these two
possibilities separately.

\subsection{Behaviour near the pole}

As $\tau \to \tau_1$, we have $s_1 \approx 1/(\tau - \tau_1) \to \infty$, $x \to 2/3$,
and, assuming $\alpha\Lambda_0 \ll 1$ (small modification), $\Lambda \to \Lambda(\tau_1)
\approx \Lambda_0$. Solution (\ref{Xp1}) behaves as in GR and the modification is
negligible. This is a special point of the solution, with all $X_i$ changing sign there
(and one of them, $X_1$, having a simple pole). All metric components are finite and
non-zero there.

\subsection{Behaviour near the would-be singularity}

Consider now the critical point.  As $\tau \to \tau_c$ from above, we have $s_1 \to 0$
and $x \to \infty$, so that $\Lambda \approx \sqrt{2 \Lambda_0 / \alpha x} \to 0$.  We
also have $x \approx s_2 / s_1^2$ with $s_2$ being finite at this point.  Therefore, the
denominator in (\ref{Xp1}) behaves as
\beq
s_1^2 ( 1+\sqrt{2\alpha \Lambda_0 x}) \approx - s_1 \sqrt{2\alpha\Lambda_0 s_2}\, ,
\eeq
where we have taken into account that $s_1 < 0$ as we approach $\tau_c$ from above. The
numerator in (\ref{Xp1}) behaves as $s_1 T^{-2}$, with small correction of order
$\sqrt{\alpha \Lambda_0}$ that can be neglected. Solution (\ref{Xp1}) then behaves as
\beq\label{Xi-mod}
X_i \approx - \frac{1}{\sqrt{2\alpha\Lambda_0 s_2} (\tau-\tau_i)^2} \sim -
\frac{1}{\sqrt{\alpha\Lambda_0}} \, ,
\eeq
which is large by absolute value (under the assumption $\alpha\Lambda_0\ll 1$) but
finite.

\subsection{Behaviour `inside' the would-be singularity}
\label{sec:would-be}

\begin{figure}
\centering
\includegraphics[width=\textwidth]{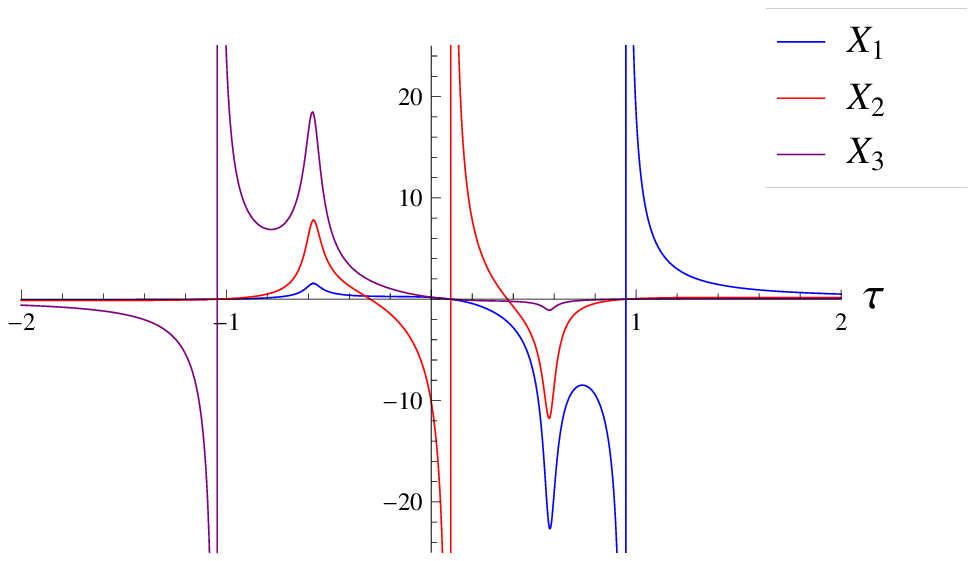}
\caption{Behaviour of the variables $X_i$ given by (\ref{Xp1}) in the modified theory
(\ref{alpha-family}) with $\alpha\Lambda_0 = 5 \times 10^{-3}$.  As in General Relativity
(see Fig.~\ref{fig:GR}), each variable $X_i$ has a pole at respective $\tau = \tau_i$.
However, the common poles at $\tau = \pm \tau_c$ disappear, and the variables $X_i$
become regular at these points.} \label{fig:modified}
\end{figure}

In the general-relativistic solution, all $X_i$ are negative and blow up near the Kasner
singularity $\tau \to \tau_c$. In the previous subsection we have seen that modification
resolves this singularity, making all $X_i$ large negative but finite. Our solution thus
smoothly continues to the region  $\tau < \tau_c$.  Since $X_i$ are finite and nonzero at
this point, $f (X)$ is also continuous.  The function $s_1$ changes sign from negative to
positive as one crosses $\tau_c$ from above. In view of the relation $f (X) = s_1 /
\Lambda$ [see (\ref{M})], this means that the sign of $\Lambda$ should become negative
below the point $\tau_c$. To ensure this, we should take the negative root of
(\ref{quadL}) in the region $|\tau| < \tau_c$\,:
\beq \label{lambda-mod}
\Lambda = \Lambda_- = - \frac{1}{\alpha x} \left( \sqrt{1 + 2 \alpha x \Lambda_0} + 1
\right) \, ,
\eeq
with $x$ defined in (\ref{x}).  Note that this branch of the solution does not have the
GR limit as $\alpha\to 0$.  Also note that the denominator in (\ref{Xp1}) is always
negative on this branch.  One can easily check that solution (\ref{Xp1}) is continuously
differentiable at the point $\tau = \tau_c$.

\begin{figure}
\centering
\includegraphics[width=.7\textwidth]{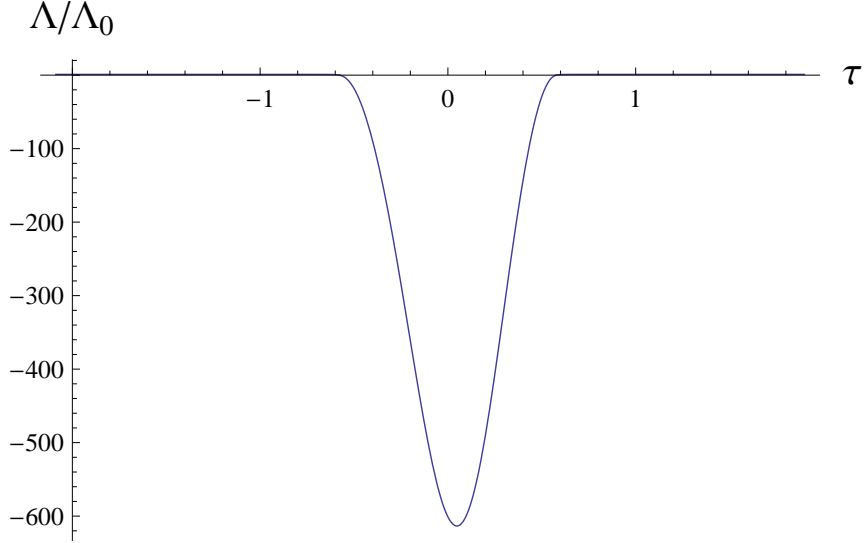}
\caption{Behaviour of $\Lambda(\tau)/\Lambda_0$ in the modified theory
(\ref{alpha-family}) with $\alpha\Lambda_0 = 5 \times 10^{-3}$.} \label{fig:lambda}
\end{figure}

As time decreases from $\tau_c$, the next special point that we encounter is $\tau =
\tau_2$. Around this point, $s_1\approx 1/(\tau - \tau_2)$, while $x \to 2/3$. Now we
have $\Lambda \to \Lambda_- (\tau_2) \approx - 3/\alpha$, which is large by absolute
value compared to $\Lambda_0$. In the neighbourhood of $\tau = \tau_2$, solution
(\ref{Xp1}) is then approximated as
\beq \label{Xp2}
X = \frac{3 T^{-3} - 2 s_1 T^{-2}}{s_1^2} \, .
\eeq
We observe that $X_1$ and $X_3$ cross zero at $\tau = \tau_2$, while $X_2 \approx 1/(\tau
- \tau_2)$ has a simple pole, approaching positive infinity as $\tau\to \tau_2$ from
above. Since all $X_i$ were negative at $\tau = \tau_c$, this means that $X_2$ crosses
zero at some $\tau_+ \in (\tau_2, \tau_c)$. It then crosses zero once again at some
$\tau_- \in (-\tau_c,\tau_2)$. This behaviour is demonstrated in Fig.~\ref{fig:modified}.
In the interval $(\tau_-, \tau_+)$, metric (\ref{canonmet}) changes signature from
$(-,+,+,+)$ to $(-,-,+,-)$, the spatial coordinate $x^2$ thus taking the role of time.
Thus, although we do not encounter singularity in the fundamental gauge field (all $\g_i$
are everywhere smooth), there is a singularity in metric (\ref{canonmet}) at the points
$\tau = \tau_\pm$, where it also changes signature.

It is interesting to plot the behaviour of the `cosmological function' $\Lambda$ for our
chosen modification. Since the non-classical branch (\ref{lambda-mod}) operates in the
time interval $[-\tau_c, \tau_c]$, we have a strongly varying function $\Lambda(\tau)$,
greatly deviating from the classical value $\Lambda_0$ in this region. This can be
clearly seen in the plot of Fig.~\ref{fig:lambda}. The limit  $\Lambda/\Lambda_0 \to 1$
as $|\tau| \to \infty$ is invisible in this plot due to resolution.

\subsection{Modified $\Lambda_0<0$ case}

In the case of GR, solution that we obtained in the time interval $(-\tau_c,\tau_c)$ was
a realisation in the theory with $\Lambda_0 < 0$. In the modified gravity theory
described above, we have obtained asymptotically De~Sitter solutions with $\Lambda_0>0$
in the regions $(\tau_c,\infty)$ and $(-\infty,-\tau_c)$ and have shown that they are
smoothly connected through a region of strong modified gravity in the interval
$(-\tau_c,\tau_c)$. At a technical level, the strongly modified solution in the region
$(-\tau_c,\tau_c)$ resulted from taking the negative root (\ref{lambda-mod}) of the
quadratic equation (\ref{quadL}) for $\Lambda(\tau)$.

In this subsection, we analyse the other possible solution. Specifically, in
$(-\tau_c,\tau_c)$ we choose the solution of the quadratic equation that has the
general-relativistic limit as $\alpha \to 0$. This is the solution that far from the
would-be singularity behaves as that in GR with $\Lambda_0 < 0$. After that, we extend
the solution beyond the interval $(-\tau_c, \tau_c)$ by taking the root
(\ref{lambda-mod}) of the equation for $\Lambda(\tau)$. In this way, we obtain regions of
strongly modified gravity in the intervals $(\tau_c,\infty)$ and $(-\infty,-\tau_c)$,
connected by an almost GR solution with $\Lambda_0 < 0$ in the interval
$(-\tau_c,\tau_c)$.

In order that the quadratic equation (\ref{quadL}) always have solution for negative
$\Lambda_0$, we now should require that $\alpha < 0$. The expression under the square
root in (\ref{lambdax}) and (\ref{lambda-mod}) will then always be positive.  As $|\tau|
\to \infty$, the function $x(\tau)\sim 2/9\tau^2$, so that $\Lambda_- (\tau) \sim -2 /
\alpha x \sim - 9\tau^2/\alpha$. We thus have a strongly varying `cosmological function'
in the regions $(\tau_c,\infty)$ and $(-\infty,-\tau_c)$.

\begin{figure}
\centering
\includegraphics[width=\textwidth]{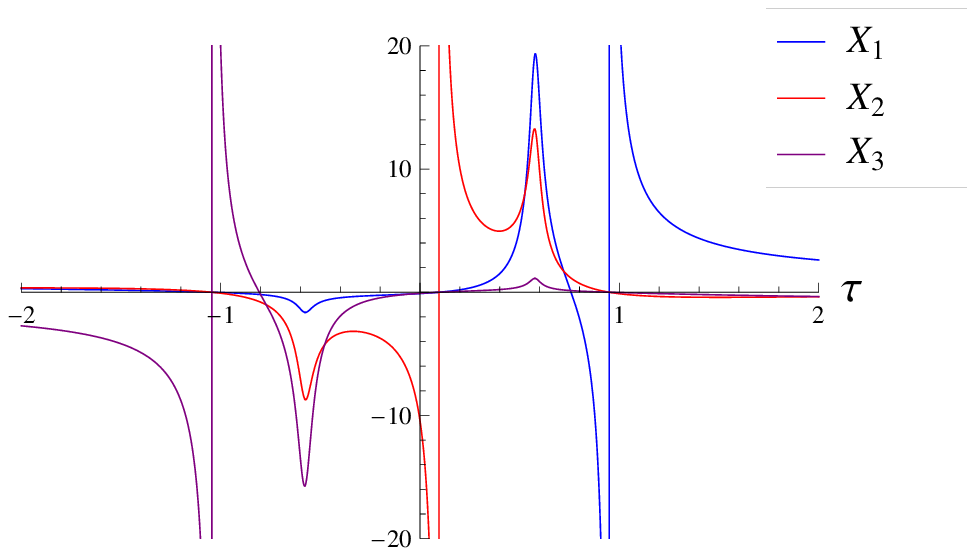}
\caption{Behaviour of the variables $X_i$ in the modified theory (\ref{alpha-family})
with $\Lambda_0 < 0$ and $\alpha\Lambda_0 = 5 \times 10^{-3}$.} \label{fig:modified-minus}
\end{figure}

It is not hard to repeat the analysis at $\tau=\pm \tau_c$ and see that there is no
singularity in this solution.  However, at some point $\tau_+ \in (\tau_c,\tau_1)$, the
variable $X_1$ will now turn to zero and change sign.  Repeating the analysis of Section
\ref{sec:would-be}, we conclude that, at this point, we will encounter singularity in
metric (\ref{canonmet}), with a signature change from $(-,+,+,+)$ to $(-,+,-,-)$.  A
similar event will take place at another point $\tau_- \in (\tau_3, \tau_c)$, where it
will be $X_3$ that will pass through zero and change sign. The signature of the metric
right below $\tau_-$ will then become $(-,-,-,+)$. Furthermore, if $\tau_2 \ne 0$, then,
depending on its sign, the variable $X_2$ will also pass through zero and change sign
either at some $\tau > \tau_1$ (for $\tau_2 > 0$) or at some $\tau < \tau_3$ (for $\tau_2
< 0$).  The signature of the metric will change accordingly at this point.
Asymptotically, by calculating the limits of solution (\ref{Xp1}) as $|\tau| \to \infty$,
we obtain
\beq \label{X-minus-infty}
X_i \to \tau_i + {\cal O} (1 / \tau) \, ,
\eeq
The behaviour of $X_i$ for $(\tau_1, \tau_2, \tau_3) = (1, 0, -1)$ is shown in
Fig.~\ref{fig:modified-minus}.

In this case, the regions $\tau \to \pm \infty$ are singular.  Indeed, integrating
(\ref{eq-g}) using (\ref{X-minus-infty}), and assuming all $\tau_i$ to be nonzero, we see
that, for large times, $\g_i \propto e^{\tau_i\tau}$. Thus, at least one of the
connection (and metric) components exponentially blows up, while another exponentially
shrinks to zero. This can be regarded as a true singularity of the solution, but
`delayed' by the modification to occur at later times.  The signature of metric
(\ref{canonmet}) in the asymptotic regions is $(-,-,-,+)$ if $\tau_2 > 0$, and
$(-,+,-,-)$ if $\tau_2 < 0$.  If $\tau_2 = 0$, then this metric has the first signature
in the region $\tau < \tau_-$, and the second signature in the region $\tau > \tau_+$.

\section{Discussion}

In this paper, we reviewed the description of the family of chiral modified
four-dimensional gravity theories. We used the language of the pure-connection
formulation, in which the modification is particularly simple to describe. The idea is
simply to consider more general functions of the curvature of the connection than the one
that produces the GR dynamics. Such a modification is guaranteed to not lead to
higher-order field equations. It may, however, result in the appearance of new
propagating degrees of freedom. The chiral family of modified gravity theories considered
in this work are known to have degrees of freedom like those present in GR --- they
continue to propagate just two polarizations of the graviton. Moreover, the argument of
\cite{Krasnov:2014eza} based on scattering amplitudes shows that this is the only family
of modified vacuum (i.e., not coupled to any matter) gravity theories in four spacetime
dimensions that do not increase the number of propagating degrees of freedom.

On the technical side, we have introduced a new parametrisation of this family of
theories. Previous studies of these chiral modified theories have used either the
first-order BF-type description, as in \cite{Krasnov:2007uu, Krasnov:2007ky,
Krasnov:2008sb, Krasnov:2010tt}, or the pure-connection description of
\cite{Krasnov:2011up, Krasnov:2011pp}. The pure-connection formalism is much easier to
deal with in practice, as the number of field components that one needs to consider is
much smaller. However, it is not free from drawbacks. One of them is the impossibility to
deal with the $\Lambda=0$ sector of GR\@. Another drawback is the presence of the square
root in Lagrangian (\ref{GR}), with its several possible branches.
Also, the field configurations where the matrix $X^{ij}$ is degenerate (across some
surfaces, for example) are problematic in the pure-connection formulation, leading to
formal singularities of $0/0$ type in the Euler--Lagrange equations. Finally, it is very
difficult to control deviations from GR in the pure-connection formulation. Indeed, it is
surprisingly difficult to describe, e.g., modification (\ref{alpha-family}) in the
pure-connection language.

All the mentioned drawbacks disappear if one allows oneself to keep, in addition to the
connection, a matrix $M^{ij}$ of auxiliary fields. One can then easily treat GR with zero
cosmological constant, as we show in the Appendix on the example of the Bianchi~I
cosmology. There is no more need to take the matrix square root $X^{ij}$ in obtaining the
GR solution. Thus, configurations where one of the eigenvalues of the matrix $X^{ij}$
turns to zero are no longer a problem, as we explicitly saw in the analysis of the GR
solution in Section \ref{sec:GR}. It is also easy to describe controlled modifications of
GR of the type (\ref{alpha-family}), as we have seen in Section \ref{sec:modified}. All
in all, we feel that `mixed' parametrisation introduced in this paper is superior to the
pure-connection one, in the sense of keeping all its advantages (it is still a connection
rather than metric formulation), while avoiding its drawbacks. This formulation has
already been put to use for other problems, see \cite{Fine:2015hef},
\cite{Krasnov:2015qhm}

The family of modifications of four-dimensional General Relativity that was described in
this paper is chiral. As a result, it is, in fact, a modification of {\em complexified\/}
GR\@. It is not difficult to state the reality conditions leading to Riemannian and split
signatures in the modified theories. However, no reality condition appropriate for the
physical Lorentzian signature is yet known, at least not in full generality. So, as
things stand, it is not clear whether the modified gravity theories described here admit
physical interpretation.

Nevertheless, there are situations where the chiral character of the theory causes no
difficulty. In this paper, we have studied one such situation. In Bianchi~I spacetimes,
the self-dual part of the Weyl curvature is real (even in Lorentzian signature). As a
result, the modifications, described here, that effectively introduce the powers of the
self-dual part of the Weyl curvature to the Lagrangian do not make the Lagrangian
complex. Because of this, the metric arising in these modified theories can be taken to
be real. In this paper, we obtained the behaviour of such real metrics in the Bianchi~I
setup by solving the evolution equations for the connection components and reading the
metric from the connection.

Our main finding is that a natural one-parameter family of modifications
(\ref{alpha-family}), with a specific choice of the sign of the parameter $\alpha$
controlling the modification, resolves the Kasner singularity of Bianchi~I spacetimes. In
the most interesting case of modified gravity with a positive cosmological constant, we
obtain a solution connecting two asymptotically De Sitter regions. This solution,
depicted in Fig.~\ref{fig:modified}, avoids the two would-be Kasner singularities of GR,
and the fundamental connection field remains regular all through the space-time.
Between the two would-be singularities, there lies a region of strong modification of
gravity. Inside this region, there are two moments of time where the metric becomes
singular, while the basic fields of the theory (the connection components), as we have
said, remain finite. This behaviour is similar to what was observed in the case of
modified black-hole solution in \cite{Krasnov:2007ky}. This type of metric singularity in
the absence of any singularity in the basic connection fields appears to be generic to
the modified theories under investigation.

The fact that modified gravity theories of this kind avoid so nicely the singularities of
GR solutions suggests that they are more than just mathematical curiosities. At the same
time, as we already noted, their physical interpretation is not clear at present. In a
more general setting, it is no longer possible to require that the metric as defined by
the connection via (\ref{urb-curv}) is real Lorentzian, as this is simply incompatible
with the dynamics. It is not yet clear how this inherent complexity of the theory should
be treated. There might exist some choice of reality conditions that reduces the dynamics
to a half-dimensional slice through the phase space with real-valued Hamiltonian and
symplectic form on this slice. If this is the case, such a real slice would provide the
sought interpretation.

A related problem is that of coupling to matter. The fundamental field of our gravity
theories is a connection, to which the matter fields should couple directly. Such
coupling in the case of GR can be obtained by the same trick of integrating out the
metric. Indeed, in the first-order formalism, all matter Lagrangians are algebraic in the
metric (i.e., do not contain derivatives of $g_{\mu\nu}$). Thus, in principle at least,
the metric can still be integrated out to produce a connection--plus--matter formulation.
In practice, however, this may be difficult. Further, it is not clear how to modify such
gravity--plus--matter theories. Alternatively, one can start with a more general family
of gauge theories as described in \cite{Krasnov:2011hi}. These describe gravity as well
as matter fields. So, some type of matter couplings can be obtained in this way. But the
problem of finding the correct reality conditions still remains in any setting --- since
matter fields couple to a complex-valued connection, some reality conditions are required
to make sense of the arising dynamics.

We end this paper by noting that the problem of reality conditions is the most pressing
one as far as classical theory is concerned. We hope that the results presented here,
even though not immediately concerned with this problem, will serve as a step towards
developing a physical interpretation of the modified gravity theories under
consideration.

\section*{Acknowledgments}

Yu.~S. is grateful to the University of Nottingham for hospitality.  Y.~H. was supported
by a grant from ENS Lyon. K.~K. was supported by ERC Starting Grant 277570-DIGT\@. The
work of Yu.~S. was supported in part by the State Fund for Fundamental Research of
Ukraine under grants F64/42-2015 and F64/45-2016.

\section*{Appendix: Chirality}

The aim of this section is to review briefly some basic facts about spinors and
self-duality in four space-time dimensions. This will clarify our usage of the term
`chirality' in this paper.

In the sense used in this paper, the notion of chirality is related to the fact that the
four-dimensional Lorentz group is doubly covered by the M\"obius group ${\rm SL}(2,\C)$.
The fundamental irreducible representations of the latter are of two different types:
2-component columns $\lambda$ on which $2\times 2$ matrices $g\in{\rm SL}(2,\C)$ act as
$\lambda\to g\lambda$, and the complex conjugate representation in terms of 2-component
columns $\bar{\lambda}$ on which the action is $\bar{\lambda}\to g^*\bar{\lambda}$, where
$g^*$ is the complex conjugate matrix. Members of these two different representation
spaces are spinors of two different types that exist in four dimensions. These
2-component spinors are chiral objects: taking the complex conjugate of a spinor of one
type, one obtains a spinor of the different type. This operation of complex conjugation
is related to the three-dimensional operation of taking the mirror image (see below),
which justifies using the terminology `chiral' also in reference to complex conjugation.

As is well-known, a 4-vector can equivalently be thought of as a bi-spinor of a mixed
type. Thus, if we refer to the spinor representation of the first type as $S_+$ and to
that of the second type as $S_-$, then the vector representation is isomorphic to
$S_+\otimes S_-$. Real elements of this space are represented by $2\times 2$ Hermitian
matrices ${\bf x}={\bf x}^\dagger$ on which the Lorentz group acts as ${\bf x} \to g{\bf
x} g^\dagger$, where $g^\dagger=(g^T)^*$ is the Hermitian conjugate of $g$. It is this
representation of 4-vectors as Hermitian matrices that provides the local isomorphism
${\rm SO}(1,3)\sim {\rm SL}(2,C)$. Taking the complex conjugate of $S_+$, one gets $S_-$.
Thus, the space $S_+\otimes S_-$ has a subspace of real, or non-chiral, objects, and
these are precisely the (real) 4-vectors.

A related chiral decomposition exists in the space of 2-forms in four dimensions. Indeed,
given a metric, we have the operation of taking the Hodge dual of a differential form.
This operation maps the space of 2-forms into itself. In the case of Lorentzian
signature, repeating this operation twice gives {\em minus\/} the identity operator:
$*^2=-1$. The eigenspaces of $*$ in the space $\Lambda^2_\C$ of complexified 2-forms are
then referred to as the spaces of self-dual $\Lambda^+$ and anti-self-dual $\Lambda^-$
2-forms. Any 2-form can be split into its self-dual and anti-self-dual parts
$\Lambda^2=\Lambda^+\oplus\Lambda^-$. Real-valued 2-forms then satisfy the condition that
their self-dual and anti-self-dual parts are the complex conjugates of each other. This
decomposition of $\Lambda^2$ is related to the spinor story we reviewed above because
$\Lambda^+$ can be shown to be isomorphic to $S_+^2$, the second symmetric power of the
fundamental representation of the first type. This is realised as the space of rank 2
spinors that are symmetric in their 2 spinor indices. Similarly $\Lambda^-\sim S_-^2$.
The operation of complex conjugation takes $S_+$ to $S_-$, and so there are real elements
in $S_+^2\oplus S_-^2$. These are real-valued 2-forms. It can also be checked that the
operation of taking the mirror image, e.g., reflecting one of the spatial coordinates,
interchanges the spaces $\Lambda^+$ and $\Lambda^-$. In this sense, the mirror reflection
is the same as complex conjugation.

\section*{Appendix: The sector of interest}

In this Appendix, we derive the diagonal ansatz used in the main text from more general
considerations.

We are interested in a general spatially homogeneous but anisotropic sector of the
theory. This is described by an ${\rm SO}(3)$ connection with components $A^i$ that are
functions of the time coordinate $\tau$ only. A general such connection is of the form
\beq
A^i = a^{ij}dx^j + b^i d \tau\, ,
\eeq
where $a^{ij}$ and $b^i$ are (complex-valued in general) functions of the time coordinate
$\tau$. Performing a time-dependent ${\rm SO}(3,\C)$ transformation, one can always make
the matrix $a^{ij}$ symmetric. After that, using $\tau$-dependent spatial translations,
one can always set the $b$-components to zero.

Thus, gauge-fixing the spatial diffeomorphisms and the gauge rotations, we are led to
consider the following sector of the theory:
\beq\label{conn}
A^i = a^{ij} dx^j \, ,
\eeq
where $a^{ij}$ is a symmetric matrix of arbitrary (complex-valued) functions of the time
coordinate only. In more mathematical language, the time evolution of our system is a
trajectory on the homogeneous group manifold
\beq
{\rm GL}(3,\C)/{\rm SO}(3,\C) \, ,
\eeq
parametrised by symmetric complex-valued $3\times 3$ matrices. We have not yet imposed
any reality conditions, this will be done below.

\subsection*{Field equations}

The field equations in the pure-connection theory are
\beq\label{feqs}
d_A \left( \frac{\partial f}{\partial X^{ij}} \right) \wedge F^j = 0\, ,
\eeq
where the matrix $X^{ij}$ is defined as in (\ref{X}) and, in view of the homogeneity of
the function $f(\cdot)$, it does not matter precisely which volume form is chosen in
(\ref{X}).

For the connection (\ref{conn}) we have:
\beq
F^i = \dot a^{ij} d\tau\wedge dx^j + \frac{1}{2} (a^{-1})^{ij} \det a \, \epsilon^{jkl}
dx^k \wedge dx^l \, ,
\eeq
where we have assumed the matrix $a^{ij}$ to be invertible,\footnote{We define the
inverse matrix by the property $a^{ik} (a^{-1})^{jk} = \delta^{ij}$.} and the dot denotes
the derivative with respect to $\tau$. Correspondingly, the matrix of the wedge products
of the curvature is given by:
\beq
F^i \wedge F^j = 2 \dot a^{(i|k|} (a^{-1})^{j)k} \det a\, d \tau \wedge dx^1 \wedge dx^2
\wedge dx^3 \, ,
\eeq
so that
\beq \label{xij}
X^{ij} \propto \dot a^{(i|k|} (a^{-1})^{j)k} \, ,
\eeq
where the proportionality means equality modulo an arbitrary function of time.

Now we can write the field equations. It is not hard to check that the $dx^i\wedge
dx^j\wedge dx^k$ component of equations (\ref{feqs}) holds automatically, in view of the
symmetry of the matrix of first derivatives of the function $f(X)$. Thus, we only need to
consider the $\epsilon^{ijk} d\tau \wedge dx^j \wedge dx^k$ part. This part reads:
\beq
\left( \frac{\partial f}{\partial X^{ij}}\right)^\cdot (a^{-1})^{jk} \det a -
\epsilon^{ipl} \epsilon^{mnk} a^{pm} \dot a^{jn} \frac{\partial f}{\partial X^{lj}} -
\epsilon^{jpl} \epsilon^{mnk} a^{pm} \dot a^{jn} \frac{\partial f}{\partial X^{li}}=0\, .
\eeq
It is convenient to multiply this equation by $a^{qk}$ and divide by $\det a$. After some
simple algebra, we get
\beq\label{eq-1}
\left( \frac{\partial f}{\partial X^{ij}}\right)^\cdot +  \frac{\partial f}{\partial
X^{ij}}\, \tr X +  \frac{\partial f}{\partial X^{jk}} (X + Y)^{ki} -  \frac{\partial
f}{\partial X^{ik}} (X + Y)^{jk} = \delta^{ij}\, \tr \left(  \frac{\partial f}{\partial
X} X\right)\, .
\eeq
Here, we have introduced the notation
\beq\label{XY}
\dot a^{ik} (a^{-1})^{jk} = X^{ij} + Y^{ij} \, ,
\eeq
where $X$ and $Y$ are the symmetric and antisymmetric parts, respectively. Now, in view
of the gauge invariance of the function $f(X)$, we have:
\beq
\frac{\partial f}{\partial X^{jk}} X^{ki} -  \frac{\partial f}{\partial X^{ik}} X^{kj}=0
\, ,
\eeq
and only the $Y$-part survives in the third and fourth terms in (\ref{eq-1}). We can also
use the homogeneity of $f(X)$ that implies
\beq
\tr \left(  \frac{\partial f}{\partial X} X\right) = f \, .
\eeq
Eventually, we get the differential equation for $X^{ij}$:
\beq\label{eqn}
\left( \frac{\partial f}{\partial X^{ij}}\right)^\cdot  +  \frac{\partial f}{\partial
X^{ij}} \tr X +  \frac{\partial f}{\partial X^{ik}} Y^{kj} +\frac{\partial f}{\partial
X^{jk}} Y^{ki}  = \delta^{ij} f(X) \, .
\eeq
Note that both the left-hand and right-hand side are explicitly symmetric in $ij$, as
they should be.

\subsection*{Reality conditions}

Let us now impose the reality conditions (\ref{reality}). For the sector of interest,
they read
\beq\label{reality-1}
\dot{a}^{ik} (\bar a^{-1})^{jk} \det \bar a + (\dot{\bar a})^{jk} (a^{-1})^{ik} \det a =0
\, .
\eeq
We will not attempt at finding the most general solution of this equation, considering
instead a particular ansatz sufficient for our purposes. Thus, we require that $a^{ij}$
are all purely imaginary:
\beq
a^{ij} = \im \g^{ij} \, , \qquad \g^{ij}\in \R \, .
\eeq
Condition (\ref{reality-1}) then boils down to
\beq\label{comm}
(\dot{\g} \g^{-1})^{[ij]}=0 \, .
\eeq
This, in particular, implies that $Y^{ij}=0$ in (\ref{eqn}).

Condition (\ref{comm}) can be stated as requiring that the matrices $\dot{\g}$ and $\g$
commute at all times. Let us consider some initial moment of time. Then we can
simultaneously diagonalise both of these symmetric matrices by an orthogonal
transformation. Then the evolution equations (\ref{eqn}) can be seen to imply that if
$\dot{\g}$ and $\g$ are diagonal at the initial moment of time, they will stay diagonal.
So, without loss of generality, we can assume $\g^{ij}$ to be diagonal at all times:
\beq
\g^{ij}={\rm diag}(\g_1,\g_2,\g_3) \, .
\eeq
The evolution equations then take the following simple form:
\ber\label{eqn1}
\left( \frac{\partial f}{\partial X^{ij}}\right)^\cdot  +  \frac{\partial f}{\partial
X^{ij}} \tr X  = \delta^{ij} f(X) \, ,  \\
X^{ij} = {\rm diag\,} \left( X_1\, , X_2 \, , X_3 \right) \, , \qquad X_i = \frac{\dot
\g_i}{\g_i} \, . \label{XY1}
\eer
It can, moreover, be assumed that all matrices appearing here are diagonal.

System (\ref{eqn1}), (\ref{XY1}) can be viewed as a system of second-order differential
equations for the functions $\g_1$, $\g_2$, and $\g_3$. However, it is more convenient to
view (\ref{eqn1}) as a system of first-order differential equations for the components of
the (diagonal) symmetric matrix $X^{ij}$. Once these are found, the components of the
connection can be found by integrating equation (\ref{XY1}). The function $f (X)$ should
be considered as given. In the main text, we study equations (\ref{eqn1}) using the
parametrisation of the function $f(X)$ in terms of $M$, as described in Section
\ref{sec:pure-conn}.

\section*{Appendix: GR solution in the physical time}

In this section, we find the GR solution of the Bianchi~I model working in the physical
time coordinate, see below. We give it here for completeness, as well as to stress the
point that the time variable used in the main text simplifies it considerably.

For the function $f (X)$ given by (\ref{GR}), which corresponds to general relativity,
equation (\ref{eq-diag}) reduces to
\beq \label{eq-y}
\left( \frac{\sum_j y_j}{y_i} \right)^\cdot  = \left( \sum_j y_j \right)^2 - \frac{\sum_j
y_j}{y_i} \sum_j y_j^2 \, ,
\eeq
where $y_j = \sqrt{X_j}$ is understood as some branch of the square root.

\subsection*{The physical time}

It is clear from (\ref{canonmet}) that the choice of physical time corresponds to the
following condition
\beq \label{casner}
\tr \sqrt{X} \det \sqrt{X} = \Lambda \, ,
\eeq
where $\Lambda$ is, in fact, the cosmological constant, which, for definiteness, we
assume to be positive. We shall denote the physical time by $t$.

\subsection*{Solution in the physical time}

Let us choose to parametrise $X_i$ as follows:
\beq \label{xparam}
X_i = \frac{\epsilon_{ijk} H_j H_k}{2 H_i} \, .
\eeq
Condition (\ref{casner}) then takes the form familiar from the Bianchi~I cosmology, in
which $H_i$ play the role of the Hubble parameters:
\beq \label{const}
H_2 H_3 + H_3 H_1 + H_1 H_2 = \Lambda \, .
\eeq
We then have
\beq
\sum_j y_j = \left( H_1 H_2 H_3 \right)^{-1/2} \, , \qquad \frac{\sum_j y_j}{y_i} =
\frac{2}{\epsilon_{ijk} H_j H_k} \, .
\eeq
Substituting these expressions into (\ref{eq-y}) and using constraint (\ref{const}), we
obtain the usual evolution equations for the Bianchi~I cosmology:
\beq\label{H-eqn}
\dot H_i + H_i^2 = \frac12 \epsilon_{ijk} H_j H_k \, .
\eeq
Solution of these equations can be parametrised by angle $\theta$ and presented in the
form
\beq
\begin{array}{l}
H_1 = \displaystyle \frac{1}{3 \sqrt{\Lambda}} \left[ \coth \sqrt{\Lambda}t + \frac{2
\sin \theta}{\sinh \sqrt{\Lambda} t} \right]
\, , \bigskip \\
H_2 = \displaystyle \frac{1}{3 \sqrt{\Lambda}} \left[ \coth \sqrt{\Lambda} t +
\frac{2 \sin \left(\theta + \frac{2 \pi}{3} \right)}{\sinh \sqrt{\Lambda} t} \right]
\, , \bigskip \\
H_3 = \displaystyle \frac{1}{3 \sqrt{\Lambda}} \left[ \coth \sqrt{\Lambda} t +
\frac{2 \sin \left(\theta - \frac{2 \pi}{3} \right)}{\sinh \sqrt{\Lambda} t} \right] \, .
\end{array}
\eeq
For $\Lambda > 0$, it describes the Kasner regime as $t \to 0$ and proceeds to the
De~Sitter space asymptotically as $t \to \infty$.  For $\Lambda < 0$, the solution is
obtained by analytic continuation, and describes evolution between two Kasner-like
singularities.

The above solution for $H$'s gives us the functions $X_1$, $X_2$, and $X_3$. We then can
integrate (\ref{eq-g}) and get the components of the connection. An equivalent, but
simpler procedure is to look for $\g$'s in the parametrisation
\beq\label{g-a}
\g_i = H_i a_i\, ,
\eeq
where $a_i$ are some yet unknown functions. Then, if we write
\beq
\frac{H_2 H_3}{H_1} = \frac{(H_1 a_1){}^{\cdot}}{H_1 a_1} = \frac{(H_2 H_3 - H_1^2)a_1 +
H_1 \dot{a_1}}{H_1 a_1}\, ,
\eeq
where we have used one of equations (\ref{H-eqn}), we see that
\beq
\frac{\dot{a_i}}{a_i} = H_i \, ,
\eeq
which allows one to solve for $a_i$ and then get the functions $\g_i$ from (\ref{g-a}).
From (\ref{metric}), we also note that the functions $a_i$ are precisely the scale
factors of the metric, i.e.,
\beq
ds^2 = - d t^2 + \sum_i a_i^2 (dx^i)^2.
\eeq
In the next section, we show how to obtain the GR solution in a simpler way by choosing a
different time coordinate.

\section*{Appendix: GR with zero cosmological constant}

In this Appendix, we find the solution of GR with zero cosmological constant using our
mixed parametrisation. The purpose of this exercise is to see that the mixed
parametrisation can be readily used to solve the equations of GR with $\Lambda = 0$,
where no pure-connection formulation is available.

For the same Bianchi~I ansatz, the evolution equations take the following form. First,
there is equation (\ref{eq-M}), which in this case takes the form
\beq\label{app-1}
\left( \Psi_i^{-1} \g \right)^{\cdot} = \tr \left( \Psi^{-1} X \right) \g = 0 \, ,
\eeq
where the zero right-hand side follows from equation (\ref{X-M-zero}).  We then get
\beq
\Psi_i = \sigma_i \g \, ,
\eeq
where $\sigma_i$ are nonzero integration constants.

Equation (\ref{X-M-zero}) and the tracelessness of $\Psi$ then imply, respectively,
\beq\label{app-4}
X_i = \zeta (\tau) \sigma_i^2 \, ,  \qquad \sum_i \sigma_i =0 \, ,
\eeq
where $\zeta(\tau)$ in the first equation is some function of time.
The second equation entails a useful relation
\beq \label{sigma}
\sum_i \sigma_i^4 = \sum_{i \ne j} \sigma_i^2 \sigma_j^2  \, .
\eeq

We look for solution of the theory in the form of power law:
\beq\label{app-g}
\g_i(\tau) = C_i \tau^{m_i} \, ,
\eeq
with some exponents $m_i$. With this ansatz for the solution, we have
\beq
X_i = \frac{\dot{\g}_i}{\g_i} = \frac{m_i}{\tau} \, .
\eeq
Equations (\ref{app-4}) give $\zeta (\tau) = C / \tau$, and imply the following relations
between $m_i$ and $\sigma_i$\,:
\beq\label{app-5}
m_i = C \sigma_i^2 \, .
\eeq
Equation (\ref{sigma}) then gives
\beq \label{app-m}
\sum_i m_i^2 = \sum_{i \ne j} m_i m_j \, .
\eeq

We now need to specify the time coordinate which, up to now, was quite arbitrary. Let us
choose $\tau$ to be the physical time coordinate $t$, in which $N^2(t) \equiv 1$.
Declaring the curvature two-forms to be self-dual, we get relations (\ref{aN}). With the
above choice for the lapse function, this gives
\beq\label{app-a}
a_1^2 = \frac{C_1^2}{m_2 m_3} t^{2(m_1+1)}\, , \quad {\rm etc} \, .
\eeq
On the other hand, according to (\ref{vol-BB}), the metric volume form (\ref{volmet})
should be a multiple of
\beq
{\rm Tr} \left( B \wedge B \right) ={\rm Tr} \left( \Psi^{-1} F \wedge \Psi^{-1} F
\right) \propto \tr (\Psi^{-2} X) \g {\cal V}_c
\eeq
[see (\ref{Xi})]. This gives
\beq
\prod_i a_i \equiv \left| \frac{\g}{\prod_i X_i} \right| \propto \left( \g t \right)^{-1}
\, ,
\eeq
which, in turn, implies $\g (t) \propto t^{-2}$, and
\beq\label{app-sum-m}
\sum_i m_i = -2 \, .
\eeq

Comparing (\ref{app-a}) to the Kasner metric (\ref{Kasner-met}) we see that we must identify
\beq
m_i = p_i-1 \, .
\eeq
Then (\ref{app-sum-m}) takes the familiar form
\beq
\sum_i p_i = 1\, ,
\eeq
and (\ref{app-m}) implies
\beq
\sum_i p_i^2 =1\, .
\eeq
We thus recover the Kasner solution.

\end{document}